\documentclass[12pt]{article}

\usepackage[english]{babel}
\usepackage{graphicx,rotating}

\begin{document}

\begin{titlepage}

\begin{center}
{\Large \bf Physics of Electroweak Gauge Bosons}

\vspace{1cm}

{\bf Klaus M\"onig}\\

DESY\\
Platanenalle 6, D-15738 Zeuthen, Germany\\
E-mail: Klaus.Moenig\@desy.de

\end{center}

\begin{abstract}
The study of gauge bosons is interesting in two respects. The properties
of gauge bosons are modified by higher order effects that are sensitive
to mass scales not directly accessible to experiment. On the other
hand interactions amongst gauge bosons are sensitive to the symmetry
structure of the theory where especially the interactions involving
longitudinally polarised bosons teach us about the mechanism of electroweak
symmetry breaking. This article explains how far a linear collider can 
increase our knowledge about the properties of and the interactions amongst
gauge bosons.
\end{abstract}
\vfill 
To appear in {\it Linear Collider Physics in the New Millennium}. Edited by K.
Fujii, D. Miller and A. Soni, World Scientific

\end{titlepage}
\newpage
\tableofcontents  


\def\CPbar{\hbox{{\rm CP}\hskip-1.80em{/}}}
\def\pd{\phantom{-}}
\def\pz{\phantom{0}}
\def\pzz{\phantom{00}}
\def\ifmath#1{\relax\ifmmode #1\else $#1$\fi}%
\def\TeV{\ifmmode {\,\mathrm{ Te\kern -0.1em V}}\else
                   \textrm{Te\kern -0.1em V}\fi}%
\def\GeV{\ifmmode {\,\mathrm{ Ge\kern -0.1em V}}\else
                   \textrm{Ge\kern -0.1em V}\fi}%
\def\MeV{\ifmmode {\,\mathrm{ Me\kern -0.1em V}}\else
                   \textrm{Me\kern -0.1em V}\fi}%
\def\keV{\ifmmode {\,\mathrm{ ke\kern -0.1em V}}\else
                   \textrm{ke\kern -0.1em V}\fi}%
\def\eV{\ifmmode  {\,\mathrm{ e\kern -0.1em V}}\else
                   \textrm{e\kern -0.1em V}\fi}%
\let\gev=\GeV
\let\mev=\MeV
\let\kev=\keV
\let\ev=\eV
\newcommand{\mco}{\multicolumn {1}{|c|}}
\newcommand{\mch}{\multicolumn {2}{|c|}}
\newcommand{\mth}{\multicolumn {3}{|c|}}
\newcommand{\mcn}{\multicolumn{1}{c}}
\newcommand{\mcl}{\multicolumn{1}{c|}}
\newcommand{\nb}{\rm{nb}}
\newcommand{\pb}{\rm{pb}}
\newcommand{\fb}{\rm{fb}}
\newcommand{\fbi}{\rm{fb}^{-1}}
\newcommand{\cm}{\mathrm{cm}}
\newcommand{\micron}{\mu\mathrm{m}}
\newcommand {\so}   {\sigma_0^{\rm{had}}}
\newcommand{\Afbpol}{A^{0,\,\ell}_{\rm {FB}}}
\newcommand{\thw}        {\theta_{\mathrm{W}}}
\newcommand{\thweff}        {\theta_{\rm{eff}}}
\newcommand{\thwefff}        {\theta_{\rm{eff}}^{\rm f}}
\newcommand{\thweffb}        {\theta_{\rm{eff}}^{\rm b}}
\newcommand{\thweffl}        {\theta_{\rm{eff}}^\ell}
\newcommand{\swsqeff}    {\sin^2\!\thweff}
\newcommand{\swsqefff}    {\sin^2\!\thwefff}
\newcommand{\swsqeffl}    {\sin^2\!\theta_{\rm{eff}}^\ell}
\newcommand{\swsqeffb}    {\sin^2\!\thweffb}
\newcommand{\Ree}      {R_{\mathrm{e}}}
\newcommand{\Rmu}      {R_{\mu}}
\newcommand{\Rtau}      {R_{\tau}}
\newcommand{\Afbze}     {A^{0,\,{\rm e}}_{\rm {FB}}}
\newcommand{\Afbzf}     {A^{0,\,{\rm f}}_{\rm {FB}}}
\newcommand{\Afbzm}     {A^{0,\,\mu}_{\rm {FB}}}
\newcommand{\Afbzt}     {A^{0,\,\tau}_{\rm {FB}}}
\newcommand{\alfas}   {\alpha_s}
\newcommand{\alfmz}   {\alfas(\MZ^2)}
\newcommand{\Rb}   {\ifmath{R_{\mathrm{b}}}}
\newcommand{\Rd}   {\ifmath{R_{\mathrm{d}}}}
\newcommand{\Rc}   {\ifmath{R_{\mathrm{c}}}}
\newcommand{\Rbz}  {\ifmath{R_{\mathrm{b}}^0}}
\newcommand{\Rcz}  {\ifmath{R_{\mathrm{c}}^0}}
\newcommand{\Gbb}   {\ifmath{\Gamma_{\mathrm{b}}}}
\newcommand{\Gcc}   {\ifmath{\Gamma_{\mathrm{c}}}}
\newcommand{\Gdd}   {\ifmath{\Gamma_{\mathrm{d}}}}
\newcommand{\Guu}   {\ifmath{\Gamma_{\mathrm{u}}}}
\newcommand{\Gss}   {\ifmath{\Gamma_{\mathrm{s}}}}
\newcommand{\MW}      {m_{\mathrm{W}}}
\newcommand{\MZ}      {m_{\mathrm{Z}}}
\newcommand{\MH}      {m_{\mathrm{H}}}
\newcommand{\MT}      {m_{\mathrm{t}}}
\newcommand{\GZ}      {\Gamma_{\mathrm{Z}}}
\newcommand{\Afb}     {A_{\mathrm{FB}}}
\newcommand{\Afbzb}    {A_{\mathrm{FB}}^{b,0}}
\newcommand{\Afbzc}    {A_{\mathrm{FB}}^{c,0}}
\newcommand{\Afbb}    {A_{\mathrm{FB}}^b}
\newcommand{\Afbc}    {A_{\mathrm{FB}}^c}
\newcommand{\Afbf}    {A_{\mathrm{FB}}^f}
\newcommand{\Gff}        {\Gamma_{\rm {ff}}}
\newcommand{\Gee}        {\Gamma_{\rm {ee}}}
\newcommand{\Gmumu}      {\Gamma_{\mu\mu}}
\newcommand{\Gtautau}    {\Gamma_{\tau\tau}}
\newcommand{\Ginv}       {\Gamma_{\mathrm{inv}}}
\newcommand{\Ghad}       {\Gamma_{\mathrm{had}}}
\newcommand{\Gnew}       {\Gamma_{\mathrm{new}}}
\newcommand{\Gnu}        {\Gamma_{\nu}}
\newcommand{\Gll}        {\Gamma_{\ell}}
\newcommand{\Gtot}       {\Gamma_{\mathrm{tot}}}
\newcommand{\GF}         {G_{\mathrm{F}}}
\newcommand{\ALR}{\mbox{$A_{\rm {LR}}$}}
\newcommand{\cAe} {\mbox{$\cal A_{\rm e}$}}
\newcommand{\cAt} {\mbox{$\cal A_{\tau}$}}
\newcommand{\cAf} {\mbox{$\cal A_{\rm f}$}}
\newcommand{\cAq} {\mbox{$\cal A_{\rm q}$}}
\newcommand{\cAl} {\mbox{$\cal A_{\ell}$}}
\newcommand{\cAb} {\mbox{$\cal A_{\rm b}$}}
\newcommand{\cAc} {\mbox{$\cal A_{\rm c}$}}
\newcommand{\cAtau} {\mbox{$\cal A_{\tau}$}}
\newcommand{\qq}{\mathrm{q}\overline{\mathrm{q}}}
\newcommand{\ff}{\mathrm{f}\overline{\mathrm{f}}}
\newcommand{\bb}{\mathrm{b}\overline{\mathrm{b}}}
\newcommand{\cc}{\mathrm{c}\overline{\mathrm{c}}}
\newcommand{\ee}{\mathrm{e}^+\mathrm{e}^-}
\newcommand{\WW}{\mathrm{W}^+\mathrm{W}^-}
\newcommand{\pp}{\mathrm{p}\bar{\mathrm{p}}}
\newcommand{\mumu}{\mu^+\mu^-}
\newcommand{\tautau}{\tau^+\tau^-}
\newcommand{\gahatf}{g_{{A}{\rm f}}}
\newcommand{\gvhatf}{g_{{V}{\rm f}}}
\newcommand{\gve}        {v_{\rm{e}}}
\newcommand{\gae}        {a_{\rm{e}}}
\newcommand{\gvf}        {v_{\rm{f}}}
\newcommand{\gaf}        {a_{\rm{f}}}
\newcommand{\ppl}  {{\cal P}_{\rm{e}^+}}
\newcommand{\pmi}  {{\cal P}_{\rm{e}^-}}
\newcommand{\ppm}  {{\cal P}_{\rm{e}^\pm}}
\newcommand{\peff}  {{\cal P}_{\rm{eff}}}
\newcommand{\pol}  {{\cal P}}
\newcommand{\Cdgz}{\ensuremath{\Delta g^\mathrm{Z}_1}}
\newcommand{\Cdgg}{\ensuremath{\Delta g^\mathrm{\gamma}_1}}
\newcommand{\Cdkz}{\ensuremath{\Delta \kappa_\mathrm{Z}}}
\newcommand{\Cdkg}{\ensuremath{\Delta \kappa_{\gamma}}}
\newcommand{\Ckg}{\ensuremath{\kappa_{\gamma}}}
\newcommand{\Ckz}{\ensuremath{\kappa_{Z}}}
\newcommand{\Clg}{\ensuremath{\lambda_{\gamma}}}
\newcommand{\Clz}{\ensuremath{\lambda_{Z}}}
\newcommand{\Cgv}[1]{\ensuremath{g^V_{#1}}}
\newcommand{\Cgz}[1]{\ensuremath{g^Z_{#1}}}
\newcommand{\Cgg}[1]{\ensuremath{g^{\gamma}_{#1}}}
\newcommand{\Ckzt}{\ensuremath{\tilde{\kappa}_Z}}
\newcommand{\Clzt}{\ensuremath{\tilde{\lambda}_Z}}
\newcommand{\Ckgt}{\ensuremath{\tilde{\kappa}_{\gamma}}}
\newcommand{\Clgt}{\ensuremath{\tilde{\lambda}_{\gamma}}}
\newcommand{\phmi}{\phantom{-}}
\setlength{\unitlength}{1cm}
\newpage

\section{Introduction}
Within any local gauge theory the properties of the gauge bosons are 
uniquely defined by the structure of the gauge group. It is thus of
special importance to study the physics of the gauge bosons. 
In the Standard Model of electroweak interactions there exist,
for the unbroken $SU(2) \times U(1)$, a triplet 
$(W^+,W^0,W^-)$, coupling to the weak isospin, and a singlet
$B$, coupling to hypercharge. All bosons at this stage are massless 
vector-particles. Due to the mechanism of spontaneous symmetry breaking 
the two neutral particles mix with a mixing angle $\theta_W$ and the 
bosons acquire mass yielding a
massive pair of charged gauge bosons, ($W^+,W^-$), a massive neutral
boson, ($Z=\cos \theta_W W^0 - \sin \theta_W B$), and the massless photon,
($\gamma=\sin \theta_W W^0 + \cos \theta_W B$).

Once the gauge group is given the gauge sector is defined by three free 
parameters,
the coupling constants for the $SU(2)$ and the $U(1)$, $g,\,g'$ and 
the vacuum expectation value of the Higgs field, $v$.

The requirement that the photon is a massless particle with vector couplings
to the fermions defines the mixing angle to be 
$g \sin \theta_W = g' \cos \theta_W = e$.
Since the couplings of the Higgs boson are pure gauge couplings the
ratio of the Z- and W-mass is given by $\cos \theta_W = \MW/\MZ$. This
relation is actually valid as long as the symmetry breaking is only
induced by doublet Higgs fields.
As another consequence of the W-Higgs couplings being pure gauge
couplings, the Fermi constant in muon decays depends only on the vacuum
expectation value, $\GF = 1/\sqrt{2}v^2$.

All formulae given above are valid on Born level. Including higher
orders in perturbation theory they receive process dependent corrections.
In these loop corrections all parameters of the model enter and also
new physics which might not be visible directly can produce measurable
effects in the radiative corrections. To test the validity of a given
model one has thus to measure a redundant set of parameters with the
best possible precision.
From measurements at lower energy the electromagnetic fine structure
constant $\alpha$ and the Fermi constant $\GF$ are known very 
accurately \cite{ref:pdg}.
Since the Z can be produced singly in $\ee$-collisions and the energy
of an $\ee$-storage ring like LEP can be calibrated with very good
precision, the Z-mass ($\MZ$) is usually taken as the third parameter
to fix the model.
It turns out then that the quantities with the highest sensitivity to
interesting radiative corrections are 
the vector and axial-vector couplings of the Z to fermions 
($\gvhatf,\,\gahatf$) and
the mass of the W.

All these quantities have been measured already with good precision at
LEP, SLD, and the {\sc Tevatron} \cite{ref:ewppe} and, as an example, the
mass of the Higgs can be constrained to be less than about 200 GeV
within the Standard Model of electroweak interactions.
However, it turns out that all these quantities can be measured
significantly more accurately at a linear collider allowing much more
stringent tests of the Standard Model.

Another feature of a local gauge theory is that the gauge bosons have
to interact amongst each other if the gauge group is non-Abelian. The
structure of these interactions is completely given by the gauge group.
The Standard Model predicts interactions between $\rm{W^+W^-Z}$ and
$\rm{W^+W^-}\gamma$ but no interactions amongst the neutral gauge bosons.

In case of a light Higgs boson electroweak interactions remain weak at
high energies and the triple gauge couplings are only modified by
loop effects. If no light Higgs boson exists electroweak interactions
become strong at the TeV scale. One therefore expects new Born level
contributions in the Lagrangian. In the latter case the process 
$VV \rightarrow V'V' \ (V,V'=W,Z)$ for longitudinally polarised vector
bosons violates unitarity at $\sqrt{s} \sim 1.5 \TeV$. 
In the Standard Model the cross section gets regularised by
the Higgs contribution. At an $\ee$-linear collider this process is
accessible with the sort of diagram shown in figure \ref{fig:vvscatfeyn}.
However since the $VV$-centre of mass energy is usually much lower than
the $\ee$-energy this process starts only to get interesting at 
$\sqrt{s}\sim 1 \TeV$.

\begin{figure}[htb]
\begin{center}
\includegraphics[height=4cm]{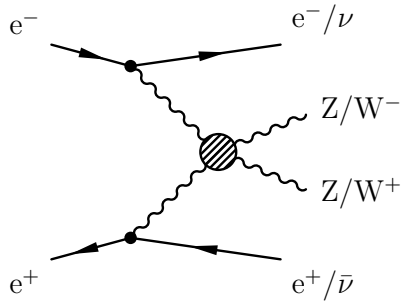}
\end{center}
\caption{Generic Feynman graph for vector-boson scattering in 
$\ee$-annihilation}
\label{fig:vvscatfeyn}
\end{figure}

In a theory of strongly interacting gauge bosons the longitudinal
degrees of freedom of the gauge bosons play the same role as the pions
in QCD. In many theories also resonances corresponding to the $\rho$, 
visible in VV-scattering exist.
In the same way, as for instance the $\rho$ is seen in $\ee
\rightarrow \pi^+ \pi^-$ also the pair production of longitudinal Ws
is modified. Such an effect will be visible in a modification of the triple
gauge couplings.

With a linear collider the physics outlined above can be tested in
several ways \cite{tdr_phys,orange,acfa}. 
In the $\ee$ mode, running on top of the Z-resonance the
precision measurements done already at LEP and SLC can be repeated
with much higher statistics and running close to the W-pair production
threshold the W-mass can be measured with high precision. 
At high energies the triple gauge couplings can be tested as
well as, at the highest energies, the strength of the vector boson
scattering. In all cases the availability of polarised beams helps a
lot and is completely indispensable for Z-pole running. In case
indications for anomalous triple gauge-couplings are found, W-pair production
in $\gamma\gamma$ collisions and single W production in e$\gamma$ collisions,
which are both sensitive to the $\rm{WW}\gamma$ vertex only help to
understand the nature of the new couplings.
Within the strongly interacting scenario the process 
$\rm{W}^-\rm{W}^- \rightarrow \rm{W}^- \rm{W}^-$ is accessible in 
$\rm{e}^- \rm{e}^-$ collisions helping to disentangle the different
models there.

\section{Production of Gauge Bosons}

Z-bosons
can be produced singly in the s-channel 
(see figure \ref{fig:zfeyn}).
For $\sqrt{s} \approx \MZ$ this process is resonant and large samples
can be produced. The total visible cross section is about $35 \nb$ so
that billions of events can be recorded with high luminosity running
at a linear collider.
\begin{figure}[htb]
\begin{center}
\includegraphics[height=3cm]{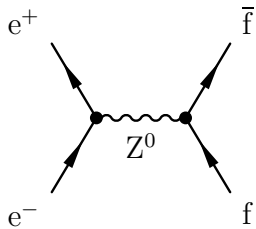}
\end{center}
\caption{Feynman graph for s-channel Z-production}
\label{fig:zfeyn}
\end{figure}

W-bosons
are produced singly 
via $\gamma$-W fusion (see figure
\ref{fig:sinwfeyn}) or in pairs 
(see figure \ref{fig:wpairfeyn}).
Since both processes are non resonant the cross sections are only a
few pb. Far above threshold, W-pair production falls like $1/s$ while
single W production rises logarithmically. Figure 
\ref{fig:wsigma} a) shows the total cross section as a function
of the centre of mass energy for both processes.
At energies around $\sqrt{s} = 500 \GeV$ they are about equal.

\begin{figure}[htb]
\begin{center}
\includegraphics[height=3cm]{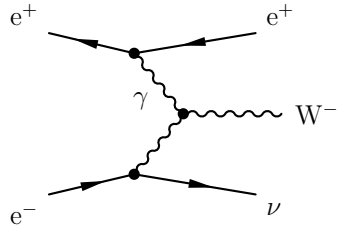}
\end{center}
\caption{Dominating Feynman graph for single W production in 
  $\ee$-annihilation}
\label{fig:sinwfeyn}
\end{figure}
\begin{figure}[htb]
\begin{center}
\includegraphics[height=3cm]{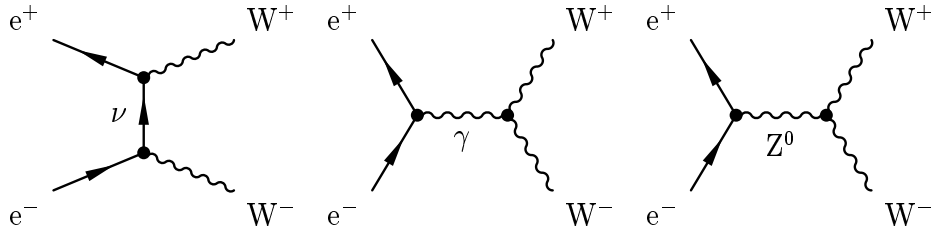}
\end{center}
\caption{Feynman graphs for the production of W-pairs in $\ee$-annihilation}
\label{fig:wpairfeyn}
\end{figure}
\begin{figure}[htb]
\begin{center}
\includegraphics[height=8.cm,bb=22 10 468 517]{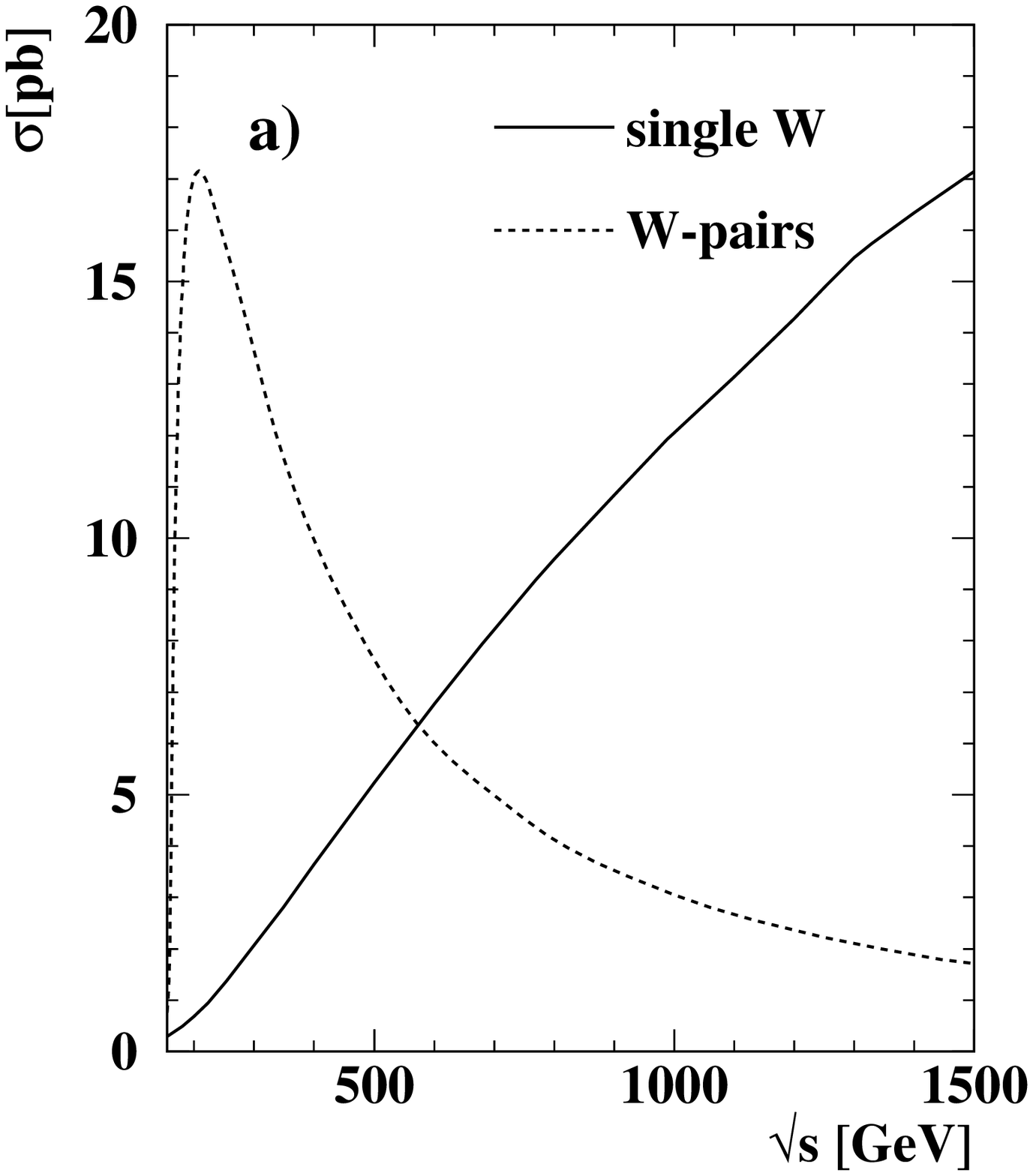}
\includegraphics[height=8.cm,bb=31 10 445 517]{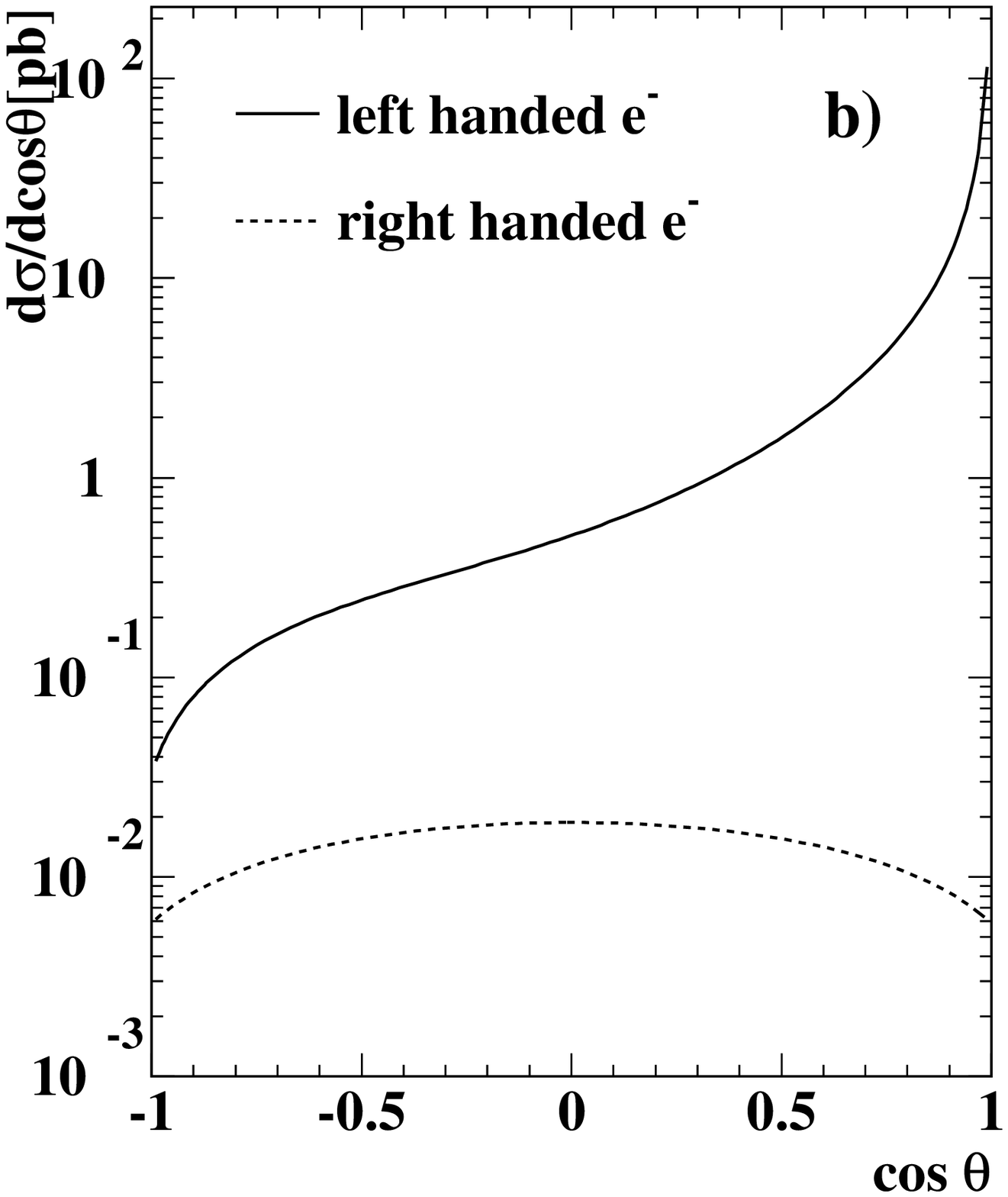}
\end{center}
\caption[]{a): total cross section for single W \cite{ref:comphep}
  and W pair production \cite{YFSWW3:2000} as
  a function of the centre of mass energy.
  b): differential cross
  section for W-pair production for different beam polarisation
  }
\label{fig:wsigma}
\end{figure}

Since the width of the W is rather large, in principle it is not
sufficient to compute the cross sections for stable Ws but instead
the full four-fermion processes have to be calculated and the
interferences between the graphs with resonant Ws and all the
background diagrams have to be taken into account. These calculations
exist and are used. At LEP-2 especially the interference between
single W production and W-pair production with one W decaying into an
electron and a neutrino turns out to be non negligible. At a linear
collider, however the average momentum of the W and the available
phase space are so large that with appropriate cuts the interferences
and the contributions from the background diagrams are almost negligible.

The radiative corrections to W-pair production are large, so that good
control of them is needed. A full one loop correction to the four fermion
process is very complicated due to the huge number of diagrams
involved and does not exist yet. However, above
threshold it is sufficient to take the full one loop corrections for
resonant Ws into account which is done in the double pole 
approximation \cite{YFSWW3:2000,ref:racoon}. The present status of the radiative
corrections is summarised in \cite{ref:lep_workshop}.

Due to the $(V-A)$ nature of the charged current couplings,
only left-handed electrons and right-handed positrons couple to
W-bosons. Single $\rm{W}^-$ production can thus be switched off or
doubled by polarising the electron beam
in the right way while single
$\rm{W}^+$ production can be modified accordingly by positron polarisation.
For the same reason the t-channel diagram for W-pair production can be
switched off by polarising one beam or enhanced by a factor of two or
four polarising one or both beams.
For energies that are much higher
than the weak boson masses, the combined~$\mathrm{Z}$
and~$\gamma$ exchange can be replaced by the neutral
member of the~$\mathrm{W}$ weak isospin triplet,
because the orthogonal combination corresponding to the
weak hypercharge boson does not couple to the
$\mathrm{W}^\pm$.  Therefore the coupling to the
electrons and positrons is also purely~$(V-A)$ at high
energies.  
Figure \ref{fig:wsigma} b) shows the differential cross section for W-pair
production for the two electron helicities at $\sqrt{s} = 500 \GeV$. 
Already at linear collider energies, the cross section
for right-handed electrons is suppressed by at least a
factor of ten relative to left-handed electrons for all
polar angles.

\section{Properties of Gauge Bosons}
\subsection{Standard Model predictions}
As already said, the gauge sector of the Standard Model of electroweak
interactions is determined by three parameters. Conveniently the three
best measured parameters are used to define the theory and the other
observables are then predicted as functions of those three. Typically
the quantities used are \cite{ref:pdg}
\begin{itemize}
\item the electromagnetic fine structure constant
at zero momentum
  transfer $\alpha$, known to a precision of $5 \cdot 10^{-8}$;
\item the Fermi constant
 in $\mu$-decays, known to $10^{-5}$;
\item the mass of the Z-boson, 
known to $2 \cdot 10^{-5}$.
\end{itemize}
On Born level all couplings and the mass of the W can then be
predicted in terms of these parameters and the partial and total
widths of the W and Z can be derived from the couplings including
small corrections due to fermion masses and due to the Kobayashi
Maskawa Matrix. The consistency of the model on this level is known
since long and the present interest is to test the theory at higher
orders in perturbation theory. If loop corrections are included the
predictions get sensitive to all parameters of the theory and also the
properties of particles, that are too heavy to be produced directly,
like the top-quark or the Higgs boson, modify the predictions. In a
similar way also parameters of a theory beyond the Standard Model can
alter these predictions, especially if the structure of the theory is
such that the parameters connected to the new heavy particles do not decouple.
All observables that can be measured with high precision can be
expressed with three types of parameters, 
the mass of the W ($\MW$) and the 
vector- and axial-vector-coupling of a fermion, f, to the Z
($\gvf,\, \gaf$). 
On Born level one has $\cos \theta_W = \MW/\MZ$,
$\gaf = 1$ and $\gvf/\gaf = 1 - 4 Q_f \sin^2 \theta_W$.

The partial width of the Z 
decaying into $\ff $ is given by
\[
\Gff \, = \, \frac{N_C^{\rm{f}} \GF \MZ^3}{6 \pi \sqrt{2}}
\left( \gvf^2 + \gaf^2 \right),
\]
where the colour-factor $N_C^{\rm{f}}$ is 1 for leptons and 3 for quarks.
Apart from corrections due
to photon exchange, $\gamma-Z$-interference and initial state
radiation,
the total cross section close to the Z-pole can then be written as
\[
\sigma_{\rm{f}} (s) \, = \, \frac{12 \pi \Gee\Gff }
{\MZ^2} \frac{s}{\left(s-\MZ^2\right)^2 + \frac{s^2}{\MZ^2}\GZ^2}
\]
so that experimentally the Z-mass and total and partial widths can be
obtained from a scan around the Z resonance. Ratios of partial widths
can be measured from cross section ratios on the peak.

$\sin^2 \theta_W$ 
can be measured with several asymmetries, 
all sensitive to
the combination $\cAf = \frac{2 \gvf \gaf}{\gvf^2+\gaf^2}$:
\begin{itemize}
\item the forward-backward-asymmetry with unpolarised beams\footnote{
    All SM predictions given here are corrected for 
    photon exchange, $\gamma-Z$-interference and initial state radiation},\\
  $\Afb^{\rm{f}} = \frac{\sigma(\cos\theta > 0)
    - \sigma(\cos\theta < 0)}{\sigma_{\rm{tot}}}
  =\frac{3}{4} \cAe {\cal A}_{\rm{f}}$;
\item the $\tau$ -polarisation and its forward-backward asymmetry for 
  unpolarised beams  $<{\cal P}_\tau> = \cAtau, \, {\cal P}_\tau^{FB} = \cAe$;
\item the left-right-asymmetry 
$\ALR = \frac{1}{{\cal P}} \frac{\sigma_L - \sigma_R}
{\sigma_L + \sigma_R} = \cAe$, independent of the final state
($\sigma_{L,R}$ denotes the cross section for left/right-handed
polarised electrons and ${\cal P}$ the beam polarisation);
\item the left-right-forward-backward-asymmetry\\

$ A_{\rm{LR,FB}}^{\rm{f}} = \frac{1}{{\cal P}} 
\frac{\left[ \sigma_L^{\rm{f}}(\cos\theta > 0) 
- \sigma_L^{\rm{f}}(\cos\theta < 0) \right]
- \left[ \sigma_R^{\rm{f}}(\cos\theta > 0) 
- \sigma_R^{\rm{f}}(\cos\theta < 0) \right] }
{\sigma_L^{\rm{f}} + \sigma_R^{\rm{f}}} =\cAf$.
\end{itemize}
The W-mass 
is measured with several reconstruction techniques in 
$\ee \rightarrow \WW$ in the continuum above threshold and in $\pp \rightarrow
\rm{W}X$ and from the cross section in $\ee \rightarrow \WW$ near threshold.

Usually the deviations from the Born level prediction are written as
\begin{eqnarray*}
  \MW^2 & = & \frac{1}{2} \MZ^2 
              \left( 1 + \sqrt{1-\frac{4 \pi \alpha}{\sqrt{2}\GF \MZ^2}
              \frac{1}{1-\Delta r}} \right)\\
  \gaf  & = & \sqrt{1 + \Delta \rho_{\rm{f}}} \\
  \gvf/\gaf & = & 1 - 4 Q_f \swsqefff,
\end{eqnarray*}
where $\Delta r$ and $\Delta \rho_{\rm{f}}$ 
are small corrections and 
$\thwefff$ is the effective weak mixing angle.

In the Standard Model and in most extensions, where new physics
appears only in loops all Z-couplings apart from the b-quark can be expressed
in terms of the Z-couplings to charged leptons with small corrections
depending only on the light fermion masses. Since the b-quark is the
isospin partner of the very heavy top-quark some
additional vertex corrections appear that are naturally enhanced by
the top mass especially if the couplings are proportional to the fermion mass,
like Higgs-couplings. In some models where new physics modifies
already the Born-level predictions, like in models where the Z and a heavy
$\rm{Z}'$ mix this need not be true.

In general there are two sorts of interesting corrections from heavy particles.
The first class of corrections stems from the mass splitting within a weak
isospin doublet. These corrections are proportional to the squared
mass difference within the doublet and led already to the successful
prediction of the top mass by LEP before its discovery.
The second class are logarithmic corrections which are non-zero also
for exact isospin symmetry.
These corrections are responsible for the present predictions on the Higgs 
mass.
The numerically largest corrections in most processes, however, are
due to the running of the electromagnetic coupling 
to the Z-scale.
$\alpha$ gets modified by about 10\% due to light fermion loops
introducing significant uncertainties in the predictions.
The
leptonic loops can be calculated reliably, however the hadronic loops
are largely affected by QCD and bear a significant uncertainty.

In addition all observables with hadronic final states have to be
corrected for virtual and real gluon radiation. As an example the
hadronic decay width of the Z has the same QCD correction as the
cross section $\ee \rightarrow \rm{hadrons}$, 
$\Ghad = \Ghad^{\rm{(no\ QCD)}}(1+\alpha_s(\MZ^2)/\pi+...)$. 

Frequently reparameterisations of the radiative correction parameters
are used where the large isospin-breaking corrections are absorbed
into one parameter, so that the others depend only on the logarithmic ones.
One example are the so called $\varepsilon$ parameters \cite{ref:epspar}
\begin{eqnarray*}
\Delta \rho_\ell & = & \varepsilon_1 \\
\swsqeffl & = & \frac{1}{2}\left(1 - \sqrt{1 - 
\frac{4 \pi \alpha(\MZ^2)}{\sqrt{2} \GF \MZ^2}} \right) 
\left( 1 -1.43 \varepsilon_1 + 1.86 \varepsilon_3 \right) \\
\frac{\MW^2}{\MZ^2} & = & \frac{1}{2}\left(1 + \sqrt{1 - 
\frac{4 \pi \alpha(\MZ^2)}{\sqrt{2} \GF \MZ^2}} \right) 
\left( 1 +1.43 \varepsilon_1 -1.00 
\varepsilon_2 - 0.86 \varepsilon_3 \right) \\
\Delta \rho_{\rm{b}} - \Delta \rho_{\rm{d}} & = & 2 \varepsilon_{\rm{b}}
\end{eqnarray*}
In this parameterisation $\varepsilon_1$ absorbs the large isospin-splitting
corrections,
$\varepsilon_3$ contains the logarithmic $\MH$ dependence while 
$\varepsilon_2$ is almost constant in the Standard Model and most extensions.
$\varepsilon_{\rm{b}}$ parameterises the additional corrections to the
Zbb vertex.

The other set of parameters that is often used are the
STU parameters \cite{ref:stdef}. 
The two parameterisations are basically
equivalent \cite{ref:pdg}. In the STU-parameters the Standard Model 
expectations are subtracted and they are multiplied by factors of order
$1/\alpha$.

\subsection{Status at present colliders}
During the first phase of LEP 
each of the four experiments has
collected about four million Z-decays at energies close to the Z-resonance and
SLD 
has recorded half a million Z-events with an average 
$\rm{e}^-$-polarisation
of about 75\%. In a second phase the LEP-experiments 
have collected
about 10000 W-pairs each, from which the W-mass has been measured and at the
{\sc Tevatron} 
about 80000 leptonic W-decays in $\pp$-collisions have been 
seen. In addition the {\sc Tevatron} has measured the top-mass to about 5 GeV
precision. A summary of all results can be found in \cite{ref:ewppe}.
Table \ref{tab:lepprec} summarises the present electroweak measurements
and compares them to the Standard Model prediction after a fit to
these data. The overall agreement is good. The two largest deviations
are $\ALR$ from SLD and $\Afbb$ from LEP. If interpreted in terms of
$\swsqeffl$, fixing $\cAb$ to its Standard Model prediction the
discrepancy between the two measurements is $2.9 \sigma$. If on the
contrary $\cAb$ is calculated from these two measurements plus the
other measurements of $\swsqeffl$ at LEP and the
left-right-forward-backward asymmetry for b-quarks at SLD $\cAb$
deviates $2.6 \sigma$ from the prediction. Since this is the largest
deviation one can construct from the data\footnote{
A $3 \sigma$ discrepancy between a $\sin^2 \theta$ in $\nu N$ scattering and 
the Standard Model prediction will be ignored here, since it is not relevant
for linear collider physics and possible theoretical uncertainties are
currently under discussion.
} 
it can well be a
statistical fluctuation.

\newcommand{\mcc}[1]{\multicolumn{1}{c|}{#1}}
\begin{table}[htbp]
  \begin{center}
\begin{tabular}{|l||r|r|r|r|}
\hline
 & \mcc{Measurement with}  &\mcc{Systematic} & \mcc{Standard} & \mcc{Pull} \\
 & \mcc{Total Error}       &\mcc{Error}      & \mcc{Model fit}&            \\
\hline
\hline
\multicolumn{5}{|l|}{
a) LEP  line-shape and lepton asymmetries: 
}\\
\hline
$\MZ$ [\GeV{}] & $91.1875\pm0.0021\pz$
                               &0.0017$\pz$ &91.1875$\pz$ & $ 0.0$ \\
$\GZ$ [\GeV{}] & $2.4952 \pm0.0023\pz$
                               &0.0012$\pz$ & 2.4961$\pz$ & $-0.4$ \\
$\so$ [nb]     & $41.540 \pm0.037\pzz$ & 0.028$\pzz$ &41.480$\pzz$ & $ 1.6$ \\
$R_\ell$       & $20.767 \pm0.025\pzz$ & 0.007$\pzz$ &20.741$\pzz$ & $ 1.0$ \\
$A^{0,\,\ell}_{\rm {FB}}$      
                & $0.0171\pm0.0010\pz  $ & 0.0003$\pz$ &0.0165$\pz$ & $ 0.7$\\
$\tau$ polarisation:                            &&&& \\
$\cAl$         & $0.1465\pm 0.0033\pz$ & 0.0010$\pz$ & 0.1483$\pz$ & $-0.6$ \\
$\qq$ charge asym.: &&&& \\
$\swsqeffl$ (${\langle Q_{\mathrm{FB}} \rangle}$)
                & $0.2324\pm0.0012\pz$ & 0.0010$\pz$ & 0.2314$\pz$ & $ 0.9$ \\

\hline
\multicolumn{5}{|l|}{
b) SLD \ 
}\\
\hline
$\cAl$ 
                 & $0.1513 \pm0.00021 $ & 0.00010     & 0.1483$\pz$ & $ 1.5$ \\
\hline
\multicolumn{5}{|l|}{
c) LEP and SLD Heavy Flavour \ 
}\\
\hline
$\Rbz{}$        & $0.21644\pm0.00065$ &  0.00053     & 0.21578     & $ 1.0$ \\
$\Rcz{}$        & $0.1718\pm0.0031\pz$ & 0.0022$\pz$ & 0.1723$\pz$ & $-0.1$ \\
$\Afbzb{}$      & $0.0995\pm0.0017\pz$ & 0.0009$\pz$ & 0.1040$\pz$ & $-2.6$ \\
$\Afbzc{}$      & $0.0713\pm0.0036\pz$ & 0.0017$\pz$ & 0.0743$\pz$ & $-0.8$ \\
$\cAb$          & $0.922\pm 0.020\pzz$ & 0.016$\pzz$ & 0.935$\pzz$ & $-0.6$ \\
$\cAc$          & $0.670\pm 0.026\pzz$ & 0.016$\pzz$ & 0.668$\pzz$ & $ 0.1$ \\
\hline
\multicolumn{5}{|l|}{
d) $\pp$ and LEP II
}\\
\hline
$\MW$ [\GeV{}] 
             & $80.449 \pm 0.034\pzz$&          &  80.394$\pzz$ & $ 1.6$ \\
$\MT$ [\GeV{}] ($\pp$)
                 & $174.3\pm 5.1\pzz\pzz$
                                        & 4.0$\pzz\pzz$
                                                      & 174.3$\pzz\pzz$
                                                                    & $ 0.0$ \\
\hline

    \end{tabular}
  \end{center}
  \caption{Results of the precision tests at LEP, SLD and the {\sc Tevatron}}
  \label{tab:lepprec}
\end{table}
As two examples, these
data can be used to constrain the Higgs mass to be less than 190 GeV
within the Standard Model or to exclude certain technicolour models
(Fig. \ref{fig:lepint}).

\begin{figure}[htbp]
  \begin{center}
    \includegraphics[height=0.47\linewidth,bb=17 37 560 565]{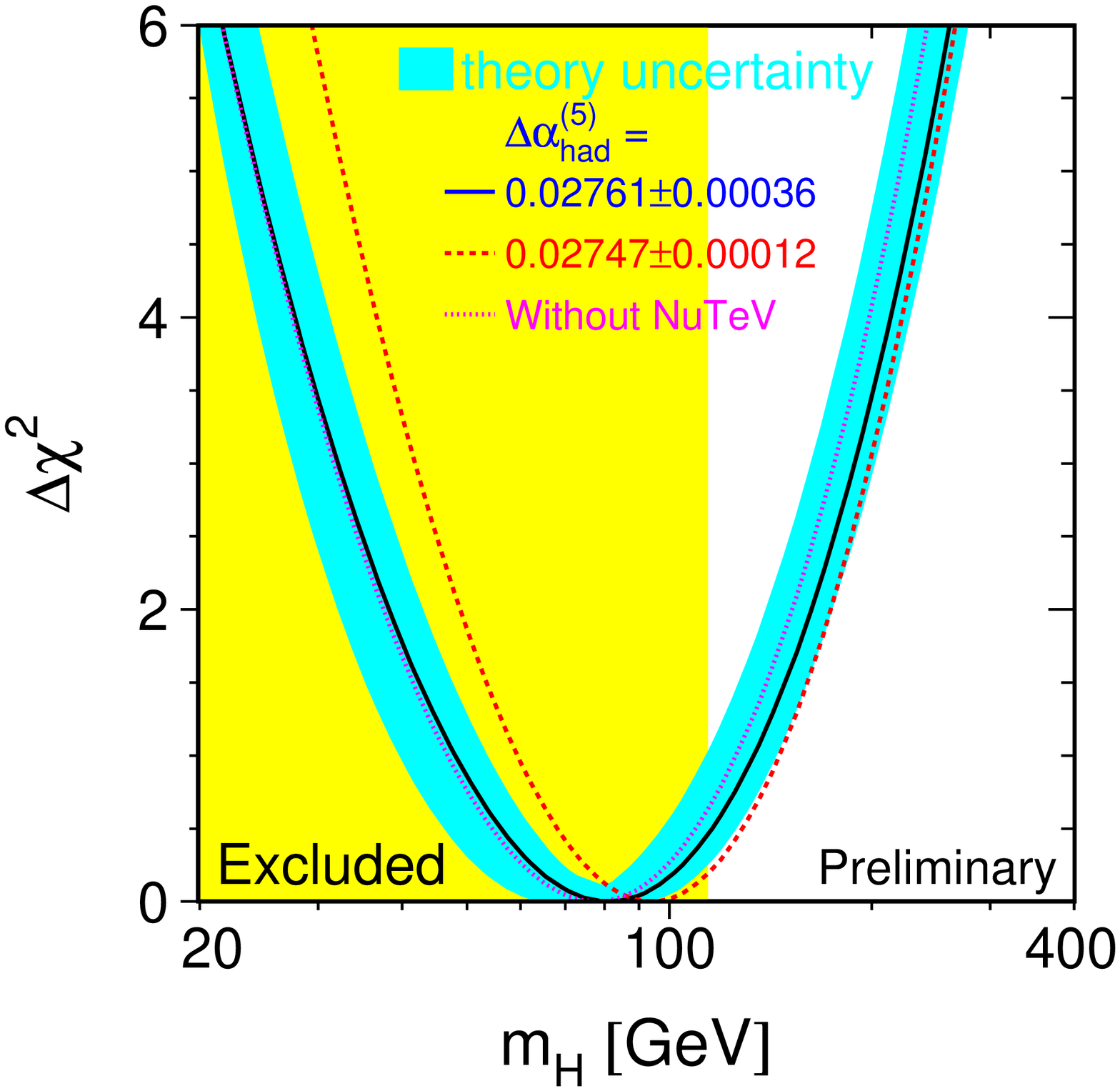}
    \includegraphics[height=0.47\linewidth,bb=5 0 500 473]{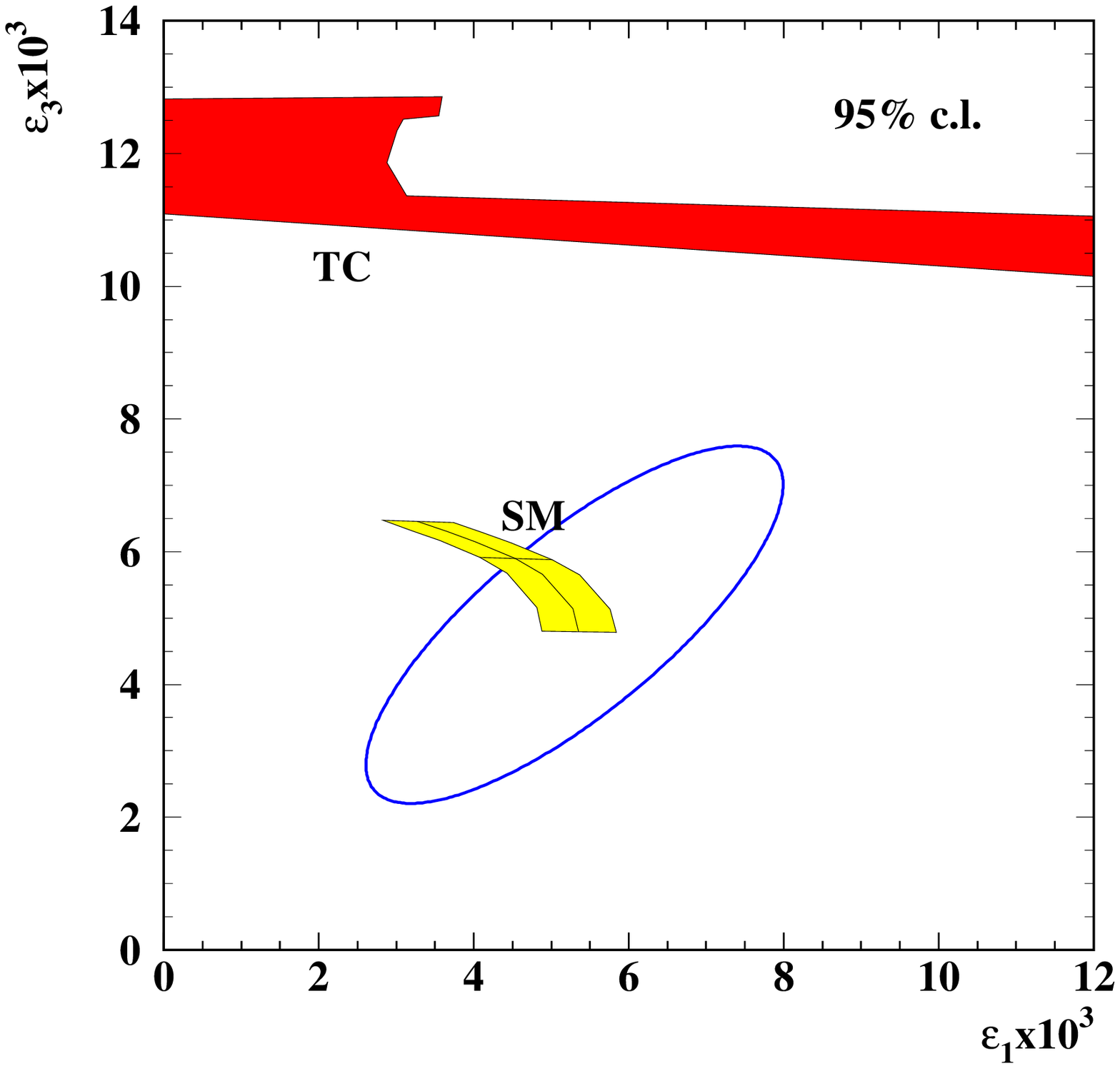}
  \end{center}
  \caption{left: $\Delta \chi^2$ of the electroweak fit as a function of the 
    Higgs mass. right: $\varepsilon$-parameters obtained from the LEP
    and SLD data compared to the Standard Model and to a technicolour model.}
  \label{fig:lepint}
\end{figure}
\subsection{Prospects for the linear collider}
A linear collider will be able to run with a luminosity of
${\cal L}\sim 5\cdot 10^{33}\rm\,cm^{-2}s^{-1}$ at energies close to
the Z-pole \cite{ref:nick}. With this luminosity it is possible to
collect $10^9$ Z-decays in about 70 days of running.
A similar luminosity should be possible close to the WW-threshold. In
addition both beams have to be polarised with polarisations of 
$\pmi=\pm 80\%$ and $\ppl=\pm (40-60)\%$. 
This corresponds to an effective polarisation of 
$\peff=\frac{\ppl+\pmi}{1+\ppl \pmi} \approx 91-95 \%$.
Additionally, the  polarisation should be switchable randomly from 
train to train.

The parameters extracted from the Z-lineshape are already today
largely systematics limited. For that reason only a modest
improvement can be expected.
Expressed in the minimally correlated observables $\MZ,\, \GZ,\,
\so = \frac{12 \pi}{\MZ^2} \frac{\Gamma_e \Ghad}{\Gamma_Z^2},\,
R_\ell = \frac{\Ghad}{\Gamma_l}$ the precision of all scan observables is 
about 0.1\%, apart from $\MZ$ which is known to
$2 \cdot 10^{-5}$.
At a linear collider it will probably not be possible to get an
absolute calibration of the beam energy 
to the same level as it was possible at LEP. 
So $\MZ$ will serve as calibration of the energy scale
and thus cannot be improved.

However a relative calibration of the energy scale to $10^{-5}$ seems
realistic, so that $\Delta \GZ = 1 \MeV$ is within reach.
With the much better detector planned in the linear collider 
designs \cite{tdr_det}
and the higher statistics to perform cross checks the systematics on
the event selection should get improved by a factor of three relative
to the best LEP experiment. This leads to a relative precision of
$\Delta R_\ell/R_\ell = 0.3 \cdot 10^{-3}$. For the $\so$ measurement
the absolute luminosity 
is needed. The experimental systematics on
this measurement might also decrease by a factor of three. However if the
theoretical uncertainty of the small angle Bhabha cross section stays
at its present value, the improvement in $\so$ will be only 30\%
leading to $\Delta \so/\so = 0.6 \cdot 10^{-3}$. To achieve these
precisions the beam-spread and the beamstrahlung need to be understood
to a few percent. This task looks difficult but possible, however
further studies are needed before a definite statement can be made.

The possible improvements in $\alpha_s(\MZ^2),\, \Delta \rho_\ell$ and
the number of light neutrino species ($N_\nu$), which can be seen as a
general parameter for extra contributions to the invisible Z width
are summarised in table \ref{tab:line}. Typical improvements are a
factor of two to three.

\begin{table}
\begin{center}
\begin{tabular}[c]{|c|c|c|}
\hline
 & LEP \protect\cite{ref:ewppe} & LC \\
\hline
$\alpha_s(\MZ^2)$ & {$ 0.1181 \pm 0.0027 $} & {$ \pm 0.0009$} \\
$\Delta \rho_\ell$ & {$ (0.50 \pm 0.10 ) \cdot 10^{-2}$} 
& {$ \pm 0.05\cdot 10^{-2}$}  \\
$N_\nu$ & {$ 2.984 \pm 0.008 $} & {$ \pm 0.004 $} \\
\hline
\end{tabular}
\end{center}
\caption{Possible improvements in the physics quantities derived from the
Z-lineshape after high luminosity Z-running at TESLA.
For $\alpha_s$ and $\Delta \rho, N_\nu=3$ is assumed.}
\label{tab:line}
\end{table}
Large improvements are expected for the asymmetries 
which in general have
low systematic errors, especially when polarised beams are
involved. From the gain in statistics and effective polarisation
relative to SLD \cite{ref:sldrev}
a factor of 50 smaller statistical error can be expected.
For $\ALR$ this corresponds to a relative error of $2 \cdot 10^{-4}$.
At present the relative error due to polarimetry is $0.5\%$ and it
cannot be expected that a polarimeter at a linear collider will be
much better than the one at SLD. The error decreases immediately by a
factor of three when positron polarisation is available with the assumed value
due to the error propagation from the single beam polarisations to the
effective polarisation, but this is still by far not enough.
However, the required precision can be obtained with the so called
Blondel scheme\cite{alain}.

In general the cross section $\ee \rightarrow \rm{Z} \rightarrow \ff$ for 
polarised beams can be written as
\begin{equation}
\sigma \, = \, \sigma_u \left[ 1 - \ppl \pmi + \ALR (\ppl - \pmi) \right].
\end{equation}

Measuring all four possible helicity combinations $\ALR$ can be
obtained without explicit knowledge of the polarisation:
\begin{equation}
  \label{eq:alrblondel}
  \ALR \, = \, \sqrt{\frac{
      ( \sigma_{++}+\sigma_{-+}-\sigma_{+-}-\sigma_{--})
      (-\sigma_{++}+\sigma_{-+}-\sigma_{+-}+\sigma_{--})}{
      ( \sigma_{++}+\sigma_{-+}+\sigma_{+-}+\sigma_{--})
      (-\sigma_{++}+\sigma_{-+}+\sigma_{+-}-\sigma_{--})}} \, ,
\end{equation}
where in $\sigma_{ij}$ $i$ denotes the sign of the positron- and $j$
the sign of the electron-polarisation. 
From an error analysis it turns out that only 10\% of the luminosity
needs to be spent on the small cross sections so that little luminosity
is lost for other measurements and the statistical error increases
only slightly due to the Blondel scheme.

In eq. (\ref{eq:alrblondel}) it is assumed that the absolute values of
the polarisation for the positive and negative helicity states are
equal. To verify this, or to get the relevant corrections,
polarimeters like the one used at SLD are still needed.
However the absolute calibration of the polarimeter analysing power,
which is the dominant error in the polarimetry, drops out in the measurement.
One possible source of uncertainty in this difference measurement
could be a difference in luminosity of the electron-laser interaction for the 
two laser polarisations. This can be overcome by using two polarimeter
channels with different analysing power.

The statistical error on $\ALR$ with $10^9$ Z-decays will be 
$\Delta \ALR = 3 \cdot 10^{-5}$. 
The variation of $\ALR$ with the centre of mass 
energy is about ${\rm d} \ALR/ {\rm d} \sqrt{s} = 2 \cdot 10^{-2}/\GeV$, so 
the beam energy 
relative to the Z-mass needs to be controlled to 
a precision of about 1\,MeV.
The amount of beamstrahlung expected in the high luminosity Z
running changes $\ALR$ by
$\Delta \ALR \, = \, 9 \cdot 10^{-4}$. It thus needs to be known to a 
precision of a few percent.
If, however, the beamstrahlung in the Z-scan to calibrate the beam energy
is the same as in the peak running it gets absorbed into an apparent
shift of the calibration constants and basically no beamstrahl
corrections for $\ALR$ are needed.

It will be thus assumed that the linear collider can measure 
$\ALR$ with a final precision of $10^{-4}$ corresponding to
$\Delta \swsqeffl= 0.000013$

$\cAb$ can be measured from the left-right-forward-backward asymmetry
with similar methods as the ones used by LEP and SLD \cite{kmrpp}.
If the lepton and the jetcharge methods from LEP are taken as a
reference a statistical error of $\Delta \cAb \sim 5 \cdot 10^{-4}$ is
possible for both methods. Not to be completely dominated by light
quark background the excellent b-tagging capabilities of the
LC-detector need to be exploited. For the jetcharge analysis a
b-tagging with $99 \%$ purity is needed, which should be feasible with
an efficiency of $75\%$. For the lepton-method a charm rejection of a
factor of 50 is needed, which can be reached with $85 \%$ efficiency.

For the lepton analysis the dominating error is then the statistical
error from $\bb$-mixing ($\Delta \cAb(\rm{mix}) \approx 9 \cdot 10^{-4}$).
This error is of purely statistical nature and cannot be reduced. For
the jetcharge method light quark background and hemisphere
correlations will both contribute to the systematic uncertainty with 
$\Delta \cAb \sim 10^{-3}$, so that, combining the two methods, a
total error of $\Delta \cAb = 10^{-3}$ seems within reach.

Another Z-observable where significant progress can be achieved is
$\Rbz = \Gamma_{\bb} / \Ghad$
which can be obtained with very small
corrections from the cross section ratio 
$\Rb = \sigma_{\bb} / \sigma_{\rm{had}}$. At LEP this quantity is
mainly limited by the understanding of the charm background, the
background from gluon splitting into $\bb$ and hemisphere correlations
due to QCD effects. These correlations arise from the energy dependence 
of the b-tag, because a gluon, emitted at a large angle, takes energy
from both jets.
With the LC-detector the charm background can be suppressed by a factor of
four relative to the LEP analyses, simultaneously reducing the
statistical error by a factor of 20. The better b-tagging and increased
statistics should also allow for a improvement in the measurement of
the gluon splitting. The energy dependence in the b-tag comes from the
large multiple scattering contribution to the impact parameter
resolution in the LEP detectors. At the LC detector the losses are
mainly due to the cut on the invariant mass of the reconstructed
particles at the secondary vertex. Since the invariant mass is a
Lorenz-invariant quantity the energy dependence and thus the hemisphere
correlations should be much smaller.
For $\Rb$ a total improvement by about a factor of five can be expected.

The W-mass 
is presently known with a precision of 34 MeV. 
However LHC 
should be able to improve its precision to about 15 MeV\cite{lhcprec}. 
The linear
collider in principle has two possibilities to measure the mass of the W,
with a scan of the W-pair-production 
threshold or reconstructing Ws at
higher energy similar to LEP.
The phase space suppression close to threshold is $\propto \beta$ for
the neutrino t-channel exchange while it is $\propto \beta^3$ for the
s-channel photon and Z-exchange. For that reason mainly the well known 
$\rm{We}\nu$-coupling enters in the prediction of the threshold
cross section and the sensitivity to anomalous triple gauge couplings
is very small. t-channel neutrino exchange is only present for
left-handed electrons and right-handed positrons. With 100\% beam
polarisation for both beams the cross section can thus be enhanced by
a factor of four for the right helicity combination and already with one
beam being polarised switched off completely. The left-right asymmetry
for the background is much smaller, so that polarisation can be used
to enhance the signal/background ratio for one polarisation state and
basically switch off the signal to measure the background for the
opposite one.

A possible W-threshold scan has been simulated for TESLA spending $100 \,\fbi$
close to $\sqrt{s}=161 \GeV$ corresponding to one year of
running \cite{ref:wscan}. 
Efficiencies and backgrounds have been assumed to be the same
as at LEP. With a total error of $0.25\%$ on the luminosity and
on the selection efficiencies $\MW$ can be measured with a total
precision of $6 \MeV$. If the efficiencies are left free in the fit,
the error increases to only $7 \MeV$ independent on the luminosity error.
The achievable errors at the scan points are compared with the
sensitivity to the W-mass in figure \ref{fig:wscan}.

\begin{figure}[htb]
\begin{center}
\includegraphics[height=8cm]{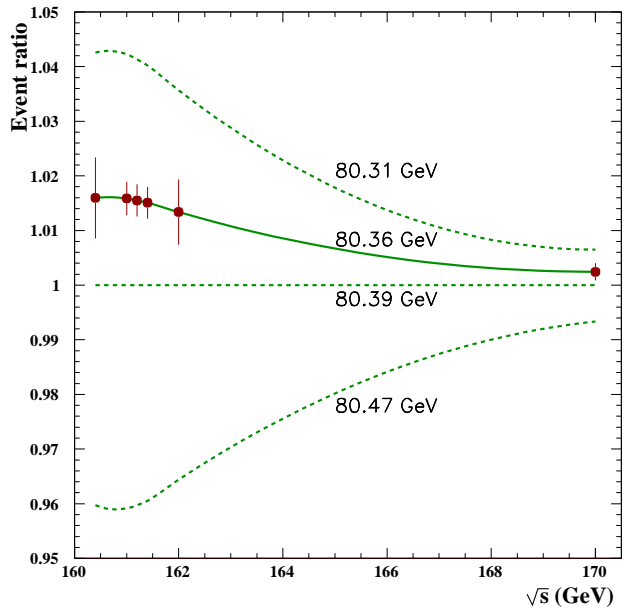}
\end{center}
\caption{Sensitivity of the W-pair threshold scan to the W-mass. The
  vertical axis shows the ratio of the cross section to the predicted
  cross section for $\MW=80.39\GeV$. The error bars represent the
  expected errors for the scan described in the text.}
\label{fig:wscan}
\end{figure}

With mass-reconstruction techniques using W-pair- and 
single-W-pro\-duc\-tion at higher energies also a statistical 
error of a few MeV
can be reached without dedicated luminosity. However it is not clear
yet, if the systematics can be brought under control.

\subsection{Interpretation of the precision measurements}

To interpret the precision measurements one has to compare them to the
model predictions at the loop level. In the theoretical prediction the
uncertainties in all input parameters enter in the uncertainty of the
prediction. The important parameters are:
\begin{itemize}
\item the running of $\alpha$ to the Z-mass. 
  $\alpha(s)$ can be expressed as 
  $\alpha(s)^{-1} = \left(1-\Delta \alpha_{\rm{lep}}(s) -
  \Delta \alpha_{\rm{had}}^{(5)}(s) - \Delta \alpha_{\rm{top}}(s)\right) \cdot
  \alpha^{-1}$. $\Delta \alpha_{\rm{lep}}(s)$ represents the leptonic
  loops which can be calculated reliably, while the ha\-dro\-nic loops,
  represented by $\Delta \alpha_{\rm{had}}^{(5)}(s)$ cause a large uncertainty.
  In principle they can be calculated from
  the cross section $\ee \rightarrow \rm{hadrons}$ for energies up to
  the Z-mass using the optical theorem. 
  However this leads to large uncertainties and everybody
  agrees to use perturbative QCD starting at some energy. However the
  energy from which on QCD can be used is under debate at 
  the moment \cite{jens}.
  If data are used until well above the resonance region
  one obtains from the low energy data 
  $\Delta \alpha_{\rm{had}}^{(5)}(\MZ^2) = (279.0 \pm 4.0) \cdot 10^{-4}$
  ~\cite{ref:aem_data,ref:aem_data_new} corresponding to an
  uncertainty in the Standard Model prediction
  of~$\swsqeffl$ of~0.00014 and 7 MeV in $\MW$.
  However,
  the sensitivity to the details of the resonance region
  can be reduced significantly, if the low energy data
  is used to fit the coefficients of a QCD operator
  product expansion instead of integrating the total cross
  section.
  If the hadronic cross section is known to 1\% up to the
  $\Upsilon$-resonances the uncertainties are $\Delta \swsqeffl = 0.000017$ 
  and $\Delta \MW = 1 \MeV$ \cite{ref:aem_data_new}.
\item A top mass error of 1\,GeV contributes an uncertainty of 0.000032 to
  the $\swsqeffl$ prediction and 6 MeV to $\MW$. 
  At TESLA it should be possible to measure $\MT$
  to $\sim 100\MeV$, so that the top contribution to both observables will be
  negligible.
\item A Z-mass 
  error of 2 MeV results in a 0.000014 uncertainty of the $\swsqeffl$
  prediction, about the same size as the experimental error and the
  uncertainty from $\alpha(\MZ)$. For $\MW$ the direct uncertainty is
  2.5 MeV. However, the linear collider calibrates for all methods the
  W-mass with the Z-mass, so that the relevant observable is
  $\MW/\MZ$. In this case the error is smaller by a factor of three.
\end{itemize}
A 10\% change in the Higgs mass 
will modify $\swsqeffl$ by $0.000031$
and $\MW$ by 6 MeV. When the Higgs is found by the time the
measurements are done its mass will be known to better than 1\% and
thus not contribute to the uncertainty in the prediction. If the Higgs
will not be found the data can be used to predict its mass to 5\% accuracy.
As an illustration figure \ref{fig:bluebandlc} compares the Higgs mass
fit of the LEP electroweak working group with the projected one after
the LC precision measurements.
\begin{figure}[htb]
\begin{center}
\includegraphics[height=8cm]{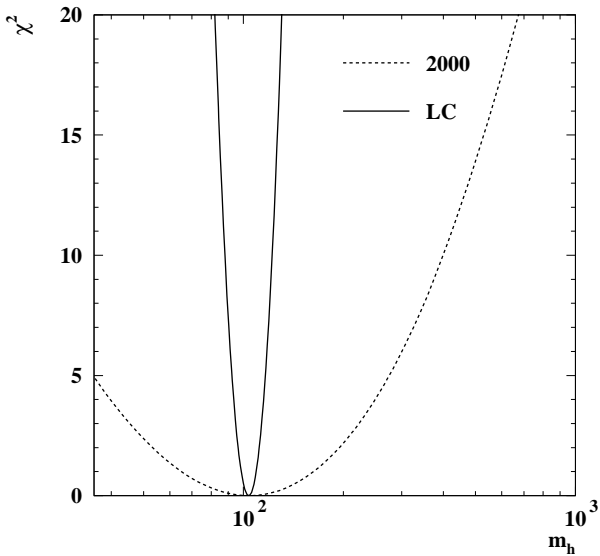}
\end{center}
\caption{$\Delta \chi^2$ as a function of 
the Higgs mass for the electroweak
precision data now and after the linear collider Z-factory running.}
\label{fig:bluebandlc} 
\end{figure}

The precision measurements can also be used to constrain extensions of
the Standard Model. As an example from the MSSM
figure \ref{fig:mwsw_susy} \cite{gigaz_th} shows the
area in $m_{\rm{A}}- \tan \beta$ that can be obtained from these data
after certain assumptions on other SUSY-parameters, mainly affecting
the stop-sector.

\begin{figure}[htb]
\begin{center}
\includegraphics[height=8cm]{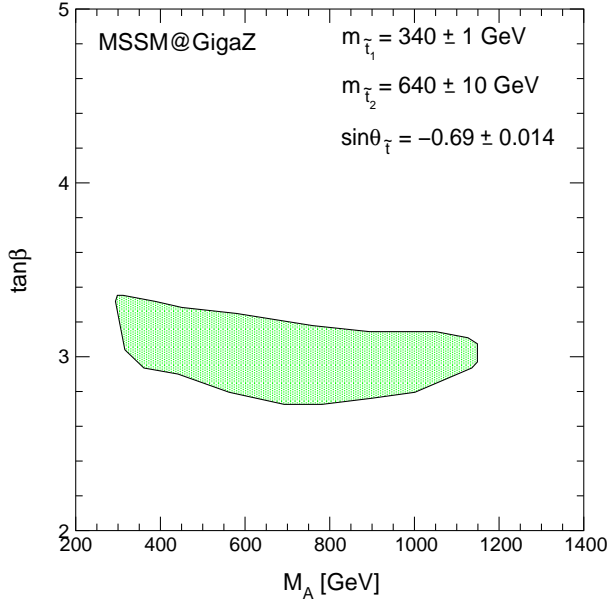}
\end{center}
\caption[]{
 The region in the $m_{\rm{A}}- \tan \beta$ plane, allowed by
$1\,\sigma$ errors by the measurements of $\MW$ and $\swsqeffl$: 
$\MW = 80.40 \gev$, 
$\swsqeffl = 0.23138$, 
and by an LC measurement of 
$\MH$: $\MH = 110 \gev$.
The other SUSY parameters including their uncertainties are given by 
$A_{\rm{b}} = -640 \pm 60 \gev$,
$\mu = 316 \pm 1 \gev$,
$M_2 = 152 \pm 2 \gev$ and
$m_{\tilde{g}} = 496 \pm 10 \gev$.
}
\label{fig:mwsw_susy} 
\end{figure}
Figure \ref{fig:eps_lc} shows the precision data in the 
$\varepsilon_1-\varepsilon_3$ plane 
for the present data and the
LC-expectation with and without $\MW$. The data are compared to the SM
expectation with $\MT=174 \GeV$ and $100\GeV < \MH < 1 \TeV$. Within
the Standard Model the Higgs is already now significantly constrained
because its trajectory is almost orthogonal to the long axis of the ellipse.
Within many extensions of the SM, however, it is easy to generate an
arbitrary $\varepsilon_1$ so that it is possible for basically any
$\MH$ to bring the prediction back into the ellipse. With the
precision of the linear collider, especially including $\MW$, the Higgs
mass will be tightly constrained without any ambiguity from $\varepsilon_1$.
\begin{figure}[htb]
\begin{center}
\includegraphics[width=0.48\linewidth,bb=0 0 544 521]{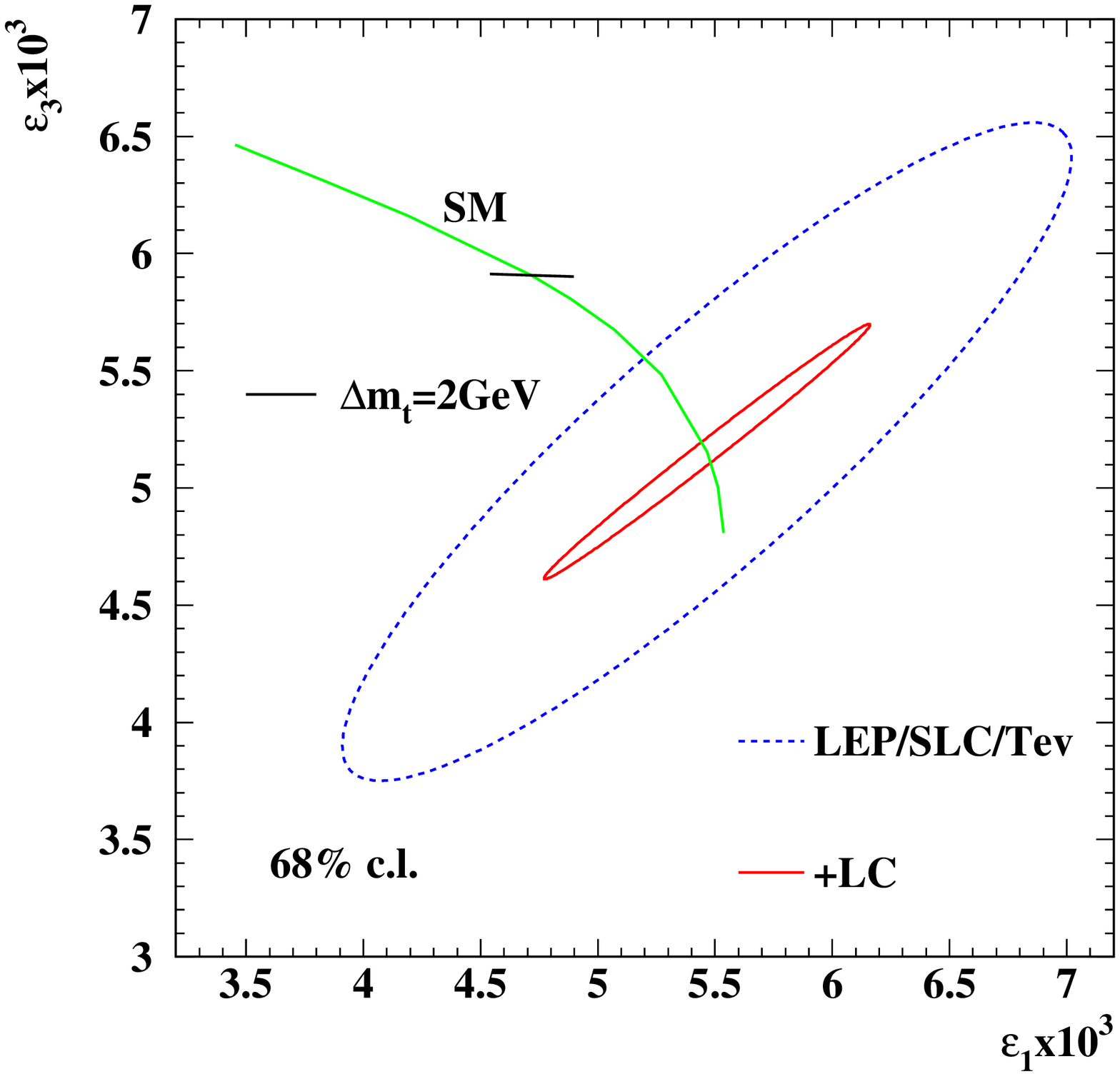}
\includegraphics[width=0.48\linewidth,bb=0 0 544 521]{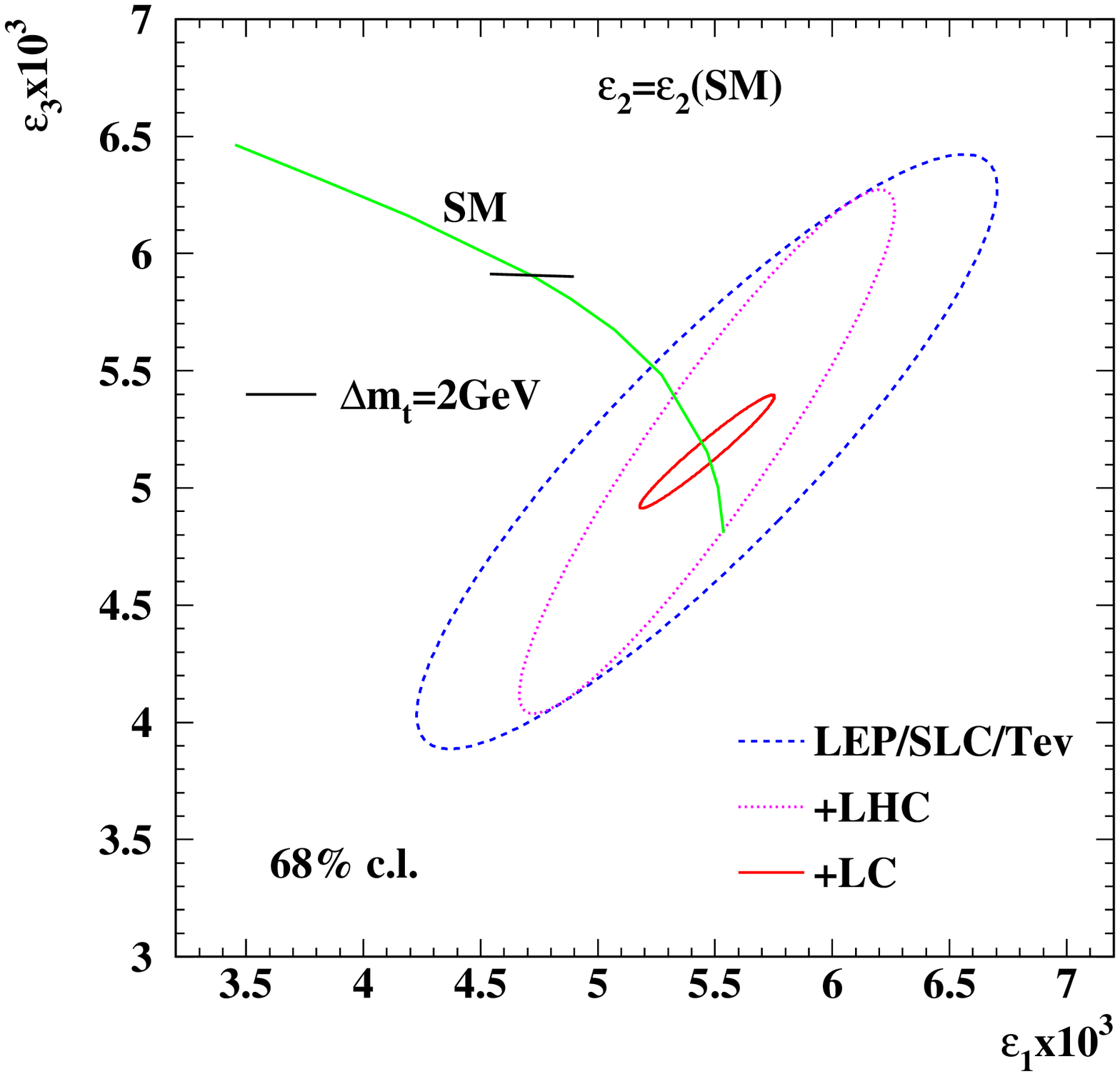}
\end{center}
\caption{
Sensitivity $\varepsilon_1-\varepsilon_3$ for the present data and the
expectation at a linear collider. 
In the left plot $\varepsilon_2$ has been left free in the fit,
equivalent to ignoring $\MW$ in the $\varepsilon_1 - \varepsilon_3$ fit.
In the right plot $\varepsilon_2$ has been fixed to its SM-value.
The line marked ``SM'' shows
the SM prediction with $\MT=174 \GeV$ and varying $\MH$ from 70 GeV
(lower end) to 1 TeV (upper end).
The effect of an uncertainty in $\MT$ of $2 \GeV$ is indicated.
$\Delta \MT=100\MeV$, expected from TESLA is inside the SM line width.
For better comparability all central values gave been assumed to be on
the SM prediction for $\MH = 100 \GeV$.
}
\label{fig:eps_lc} 
\end{figure}

As an example of an application of the model independent analysis, figure
\ref{fig:stthdm} shows in the S-T plane 
the predictions from the
2-Higgs-Doublet Model (2HDM) 
for cases, where a light Higgs exists but
cannot be seen, compared to the present data and the expectations from
a linear collider \cite{thdm}. 
Only a linear collider can distinguish between the
Standard Model with a light Higgs and the 2HDM. One can also see, that
for these sort of analyses the good precision in the W-mass 
is needed.
\begin{figure}[htbp]
\begin{center}
\includegraphics[width=\linewidth]{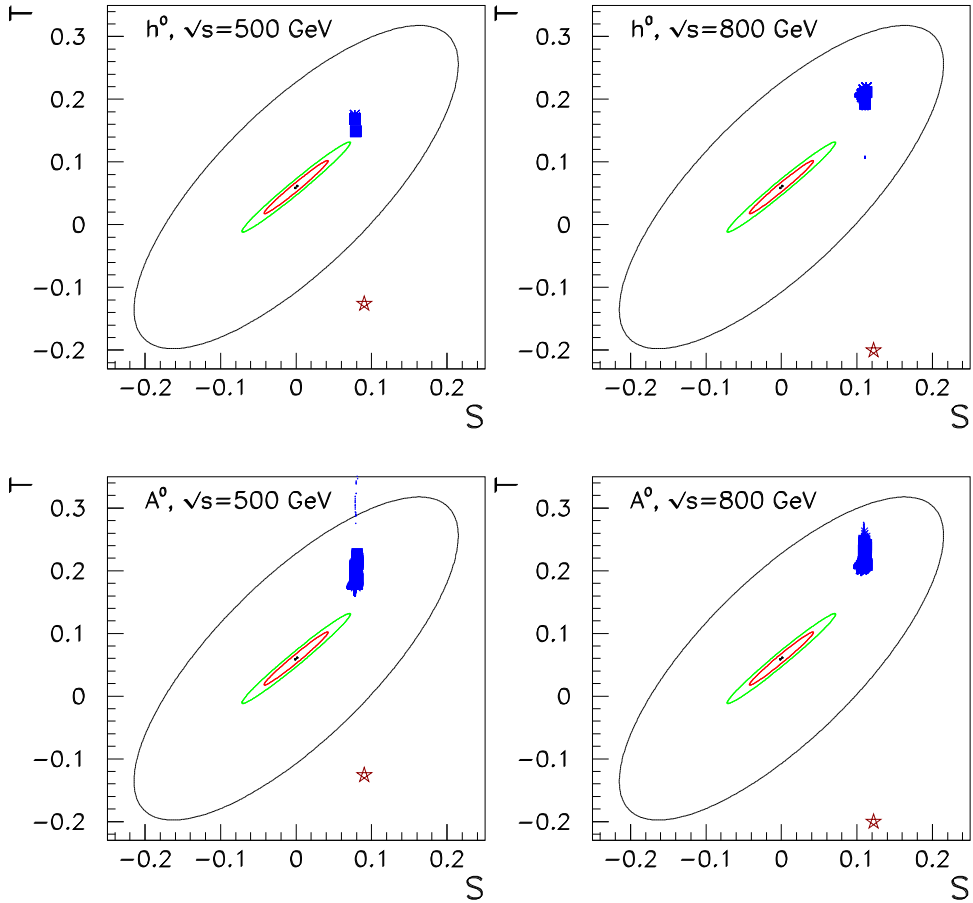}
\end{center}
\caption[bla]{Prediction for S and T from the 2 Higgs doublet model with a
  light Higgs for the cases where no Higgs is found compared to the
  current electroweak data and the projection for Giga-Z. The star
  denotes the Standard Model prediction if the Higgs mass is 
  $\sqrt{s} - 100 \GeV$.
}
\label{fig:stthdm} 
\end{figure}

In summary, the properties of gauge bosons, namely the mass of the W
and the fermion couplings of the Z will be highly sensitive to loop
corrections. They will thus tightly constrain the SM, in case no new
physics is found until then or help to measure the parameters of the
new theory, if new phenomena are seen.

\section{Measurements of the CKM matrix}
The CKM matrix connects the weak eigenstates of down type quarks
$(D_1')$ to
their mass eigenstates $(D_i)$ \cite{ref:pdg}:
\[
   D_i' \, = \, \sum_j V_{ij} D_j
\]
The matrix has to be unitary to avoid flavour changing neutral
currents on tree level, which have not been seen experimentally. For
three flavours it can be parameterised by three angles and one complex
phase, introducing CP violation.
The coupling of an up-type quark $U_i$ and a down-type quark $D_j$ to
the W-boson, normalised to the leptonic coupling of the W can thus be
written as
\[
  g(U_i,D_j,W) = V_{ij} g(\ell,\nu,W) .
\]
Traditionally the CKM matrix elements are measured in weak meson
decays. The absolute values of the elements
can be measured from corresponding decay
rates. Normally large event samples are available so that statistical
errors are very small. However the hadronic matrix elements for the
weak transitions are quite uncertain and the errors from this source
are totally dominant. The phases of the elements only show up
in interference patterns and are much harder to measure. 
A linear collider can contribute in three ways to the measurement of
the CKM matrix elements:
\begin{itemize}
\item The elements $V_{{\rm t}i}$ can be accessed in top decays. This is
  discussed in chapter 3.
\item The CP violating phases can be studied in B-decays in the Giga-Z
  running. The methods are essentially the same as at the $\ee$
  B-factories and the B-physics experiments at hadron machines, namely
  BTeV and LHCb and the expected precision is typically in between the
  lepton- and the hadron machines. A detailed discussion can be found in
  \cite{bphys_ee,bphys_had,monhaw_epj}.
\item The decay width of a W into a quark pair 
  depends on the corresponding matrix element:
  \[
  \Gamma(W^+ \rightarrow U_i \overline{D_j}) = 
  |V_{ij}|^2 \Gamma(W^+ \rightarrow \ell \overline{\nu}) 
  (1+{\cal C}_{\rm{QCD}}),
  \]
  where ${\cal C}_{\rm{QCD}}$ parameterises QCD corrections.
  This method will be described in some detail in the 
  following \cite{ref:james}.
\end{itemize}
The experimental problem in the measurement of the CKM matrix elements
from W decays is the tagging of the quark flavours. For b- and c-quarks
there exist well understood methods, tagging the decays of long lived
b- and c-flavoured hadrons with the microvertex detector. Already at
LEP and SLD 
these algorithms have been used successfully in many
analyses and they are essential in Higgs physics at a linear collider.
For light quarks no high efficiency, high purity methods
exist. However it is known that the leading hadron in a jet often
contains the primary quark so that it can be used for flavour tagging. 
The Z decay widths to b- and c-quarks and to the sum of all quarks are known 
with very good precision from LEP and SLD and they agree well with the
Standard Model prediction. It is thus reasonable to assume that also
the partial width to u-, d- and s-quarks agree with the prediction to
the same precision. The tagging purities and efficiencies for all
quarks in Z-decays using the microvertex methods for b- and c-quarks
and leading hadrons for light quarks can thus be measured accurately
in Z-running at the linear collider. 
Fragmentation properties depend only
logarithmically on the energy scale. The extrapolation from the Z-mass
scale to the W-mass scale can thus be done safely using the usual
parton shower fragmentation models. Typically the efficiencies are 2\%
larger for the Ws.
The only complication comes from the fact that, due to isospin
invariance, the probability for a primary u- or d-quark to end up in a
charged pion is about the same, making the fit to the Z-data unstable. 
However, because of charge
conservation the W always has to decay in one up- and one down-type 
(anti-)quark. If, as an example, one quark from a W is tagged as
charm the other one can be a down-quark, but not an up-quark.
So the problem can be solved if the Z- and the W-data
are fitted simultaneously with the tagging efficiencies and the
W-branching ratios as free parameters, without introducing additional
model dependence.
Table \ref{tab:udstag} shows the obtainable efficiencies for light quarks 
using $\pi^\pm,\, K^\pm, \, K^0_s,\, p$ and $\Lambda$.

\begin{table}[htbp]
  \begin{center}
    \begin{tabular}{|c|c|c|c|}
      \hline
      & d & u & s \\
      \hline
      $\pi^\pm$ & 0.209 & 0.209 & 0.130 \\
      K$^\pm$   & 0.056 & 0.074 & 0.122 \\
      p         & 0.015 & 0.025 & 0.020 \\
      K$^0_S$   & 0.007 & 0.006 & 0.019 \\
      $\Lambda$ & 0.003 & 0.003 & 0.008 \\
      \hline
    \end{tabular}
  \end{center}
  \caption{Light quark tagging efficiencies of leading hadrons with $x_p>0.2$}
  \label{tab:udstag}
\end{table}
For the analysis of the CKM-matrix all events containing W bosons can be
used. Especially attractive, however, are single W events. 
The W in
these events has on average a very small momentum so that the typical jet 
energies are close to those in Z-pole events. So in these events the
extrapolation of the detector efficiencies, which vary strongly with energy, 
especially for the dE/dx measurement, is very small.
Table \ref{tab:ckmres} shows the sensitivity of a linear collider with 
$1500\,\fbi$ integrated luminosity compared to present and expected accuracy.

\begin{table}[htbp]
  \begin{center}
\begin{tabular}{|c|c|c|c|} \hline
             & current uncertainty & projected other & LC        \\ 
\hline
 $|V_\mathrm{ud}|$  & $\pm $ 0.0008       &              & $\pm $0.0028 \\ 
 $|V_\mathrm{us}|$  & $\pm $0.0023        &              & $\pm $0.0124 \\ 
 $|V_\mathrm{ub}|$  & $\sim \pm $0.008    & $\pm $0.0004 & $\pm $0.011  \\ 
\hline
 $|V_\mathrm{cd}|$  & $\pm $0.0016        &              & $\pm $0.0072 \\
 $|V_\mathrm{cs}|$  & $\pm $0.01          &              & $\pm $0.0017 \\
 $|V_\mathrm{cb}|$  & $\pm $0.0019        & $\pm $0.0012 & $\pm $0.0011 \\
\hline
\end{tabular}
\end{center}
\caption[]{Current \cite{ref:pdg} and projected \cite{bphys_ee}
  uncertainty of the CKM matrix elements compared to the LC
  measurement from W-decays.
  }
\label{tab:ckmres}
\end{table}

A large gain can be obtained in $V_{\rm cs}$. However, if the unitarity of
the CKM matrix is imposed, the present error in this element shrinks by
more than one
order of magnitude. Also in $V_{\rm cb}$ some progress is possible. This
element is important for the interpretation of the CP-violation results
in B-decays. If it is determined in B-decays the error is totally
dominated by the theoretical uncertainty on the hadronic matrix
elements, while in the determination from W-decays the error is
statistical. The independent cross check is thus highly desirable.

\section{Interactions amongst Gauge Bosons}
The Standard Model predicts trilinear interactions between WWZ and
WW$\gamma$, however there should be no interactions between neutral
gauge bosons.

Single W production 
and W-pair production 
are sensitive to the triple
gauge-boson couplings (TGC). In W-pair production the cross sections
for neutrino t-channel exchange and $\gamma$,Z-s-channel exchange
would rise individually with energy violating unitarity at some
point. As demonstrated in figure \ref{fig:wpaircanc} this rise is
cancelled by the interference term, so that the resulting cross
section shows the usual $1/s$ behaviour. For this reason W-pair
production is extremely sensitive to the TGCs. To preserve the gauge
cancellations, however, any anomalous coupling has to vanish for 
$\sqrt{s} \rightarrow \infty$. 
For the linear collider studies using W-pairs in principle this is not a 
problem, since the couplings are measured at a well defined scale.
In the hadron collider studies the running of the couplings with the
energy is often parameterised by a form factor $x' = \frac{x}
{\left(1 + s/\Lambda^2 \right)^n}$ with $n>0.5$ or $n>1$ depending on the 
type of coupling \cite{lhcprec}. 
$\Lambda$ can be interpreted as the scale where the new physics,
responsible for the anomalous couplings sets in, so that the effective 
parameterisation becomes meaningless above this scale.

\begin{figure}[htb]
\begin{center}
\includegraphics[height=8cm,bb=0 17 567 497]{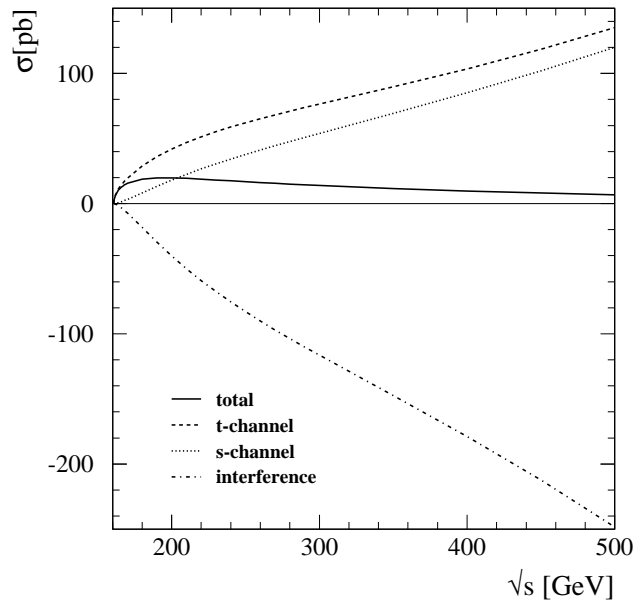}
\end{center}
\caption{W-pair production cross section for the different production 
processes separately, the interference and for the sum of all}
\label{fig:wpaircanc}
\end{figure}

Contrary to W-pair production the effective scale of single W
production is always much lower than the centre of mass energy. The
flux functions for radiating a W or a photon off the initial state
electrons, peak at low energy so that the produced W is on average
almost at rest.

In models where the gauge interactions 
remain weak they
can be parameterised with a linear Lagrangian. Keeping terms with
dimension up to six the most general Lagrangian for WWV (V=Z,$\gamma$) is
\begin{eqnarray*}
i {\cal L}_{eff}^{WWV} & = & g_{WWV} \cdot [ \\
 & & g_1^V V^\mu \left( W^-_{\mu\nu}W^{+\nu} - W^+_{\mu\nu}W^{-\nu} 
\right) + \\
 & & \kappa_V W^+_{\mu}W^-_{\nu}V^{\mu\nu} + \\
 & & \frac{\lambda_V}{m_W^2} V^{\mu\nu} W_\nu^{+\rho} W^-_{\rho\mu} + \\
 & & i g_5^V \epsilon_{\mu\nu\rho\sigma} \left(
\left(\partial^\rho W^{-\mu}\right) W^{+\nu} - 
  W^{-\mu} \left(\partial^\rho W^{+\nu}\right) \right)V^\sigma + \\
 & & i g_4^V W^-_\mu W^+_\nu 
\left( \partial^\mu V^\nu + \partial^\nu V^\mu \right) - \\
 & & \frac{\tilde{\kappa}_V}{2} W^-_\mu W^+_\nu 
\epsilon^{\mu\nu\rho\sigma} V_{\rho\sigma} - \\
 & & \frac{\tilde{\lambda}_V}{2m_W^2} W^-_{\rho\mu} W^{+\mu}_{\ \ \ \ \nu}
\epsilon^{\nu\rho\alpha\beta} V_{\alpha\beta} ]
\end{eqnarray*}
with $V = \gamma,\,Z, \ g_{WW\gamma} = e, \ g_{WWZ} = e \cot \theta_W$
and $V_{\mu\nu} = \partial_\mu V_\nu - \partial_\nu V_\mu$.

Electromagnetic gauge invariance requires $g_1^\gamma (q^2=0) = 1$ and
$g_5^\gamma (q^2=0) = 0$.
In the Standard Model one has 
$g_1^V = \kappa_V = 1$. All other couplings are equal to zero.

In term of these parameters the magnetic dipole moment of the W is
given by $\mu_W = \frac{e}{2m_W} (1+\kappa_\gamma+\lambda_\gamma)$ and the
electric quadrupole-moment by 
$q_W = - \frac{e}{m_W^2} (\kappa_\gamma-\lambda_\gamma)$.

From the terms in the effective Lagrangian the ones multiplied by
$g_1,\,\kappa$ and $\lambda$ are C- and P-conserving,
the one proportional to $g_5$ violates C and P, however is CP-conserving
and the ones with $g_4,\,\tilde{\kappa}$ and $\tilde{\lambda}$ 
are CP-violating.

It is expected that deviations from the Standard Model should show up first 
in the C,P conserving couplings. At an $\ee$-collider the  process that is
by far most sensitive to triple gauge couplings is W-pair
production. However, if no beam polarisation 
is available, it is not
possible to separate WW$\gamma$ and WWZ couplings with this
process. If one requires that the anomalous couplings are C-
and P-symmetric, preserve $SU(2) \times U(1)$ symmetry without adding 
operators of higher dimension and are not
excluded by the LEP1 precision data one is left with the three
couplings \cite{ref:leptwo_tgc}:
\begin{eqnarray}
\nonumber
\Delta g_1^Z & \ & \ \\
\Delta \kappa_\gamma & = &
- \frac{\cos^2 \theta_W}{\sin^2 \theta_W}\left( \Delta \kappa_Z - \Delta g_1^Z 
\right) \label{eq:surel} \\
\nonumber
\lambda_\gamma & = & \lambda_Z 
\end{eqnarray}
Since the initial state couplings $\ee\gamma$ and $\ee \rm{Z}$ depend
differently on the beam polarisation, polarised beams can be used to
separate WW$\gamma$ and WWZ couplings, as demonstrated in figure
\ref{fig:tgcpol_pr}. With enough statistics it is
thus possible to measure the five C,P-conserving couplings
simultaneously. The C,P-violating couplings can be separated easily
by constructing C,P-odd observables.
\begin{figure}[htb]
\begin{center}
\includegraphics[width=0.45\linewidth,bb=18 10 449 515]{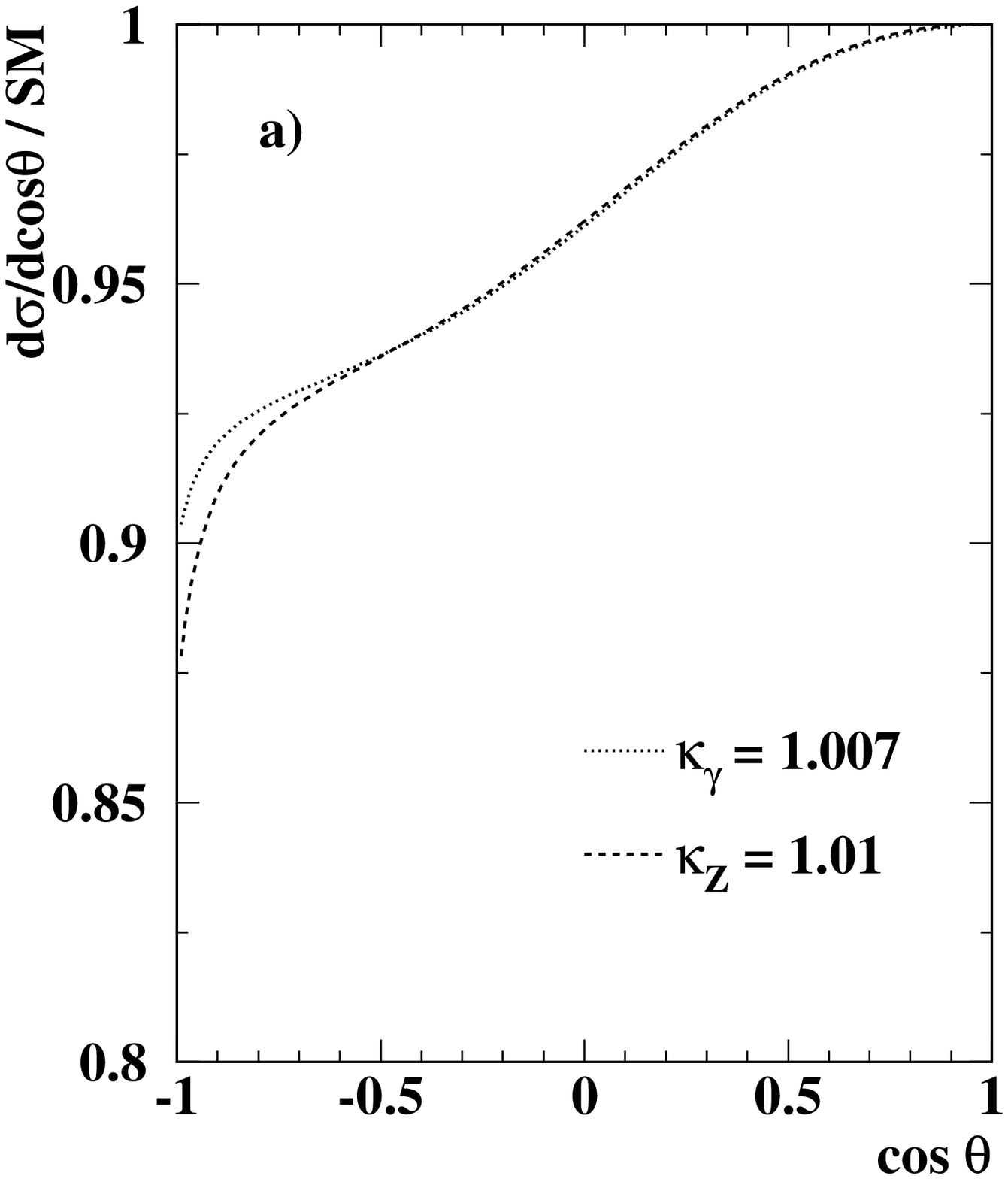}
\includegraphics[width=0.45\linewidth,bb=18 10 449 515]{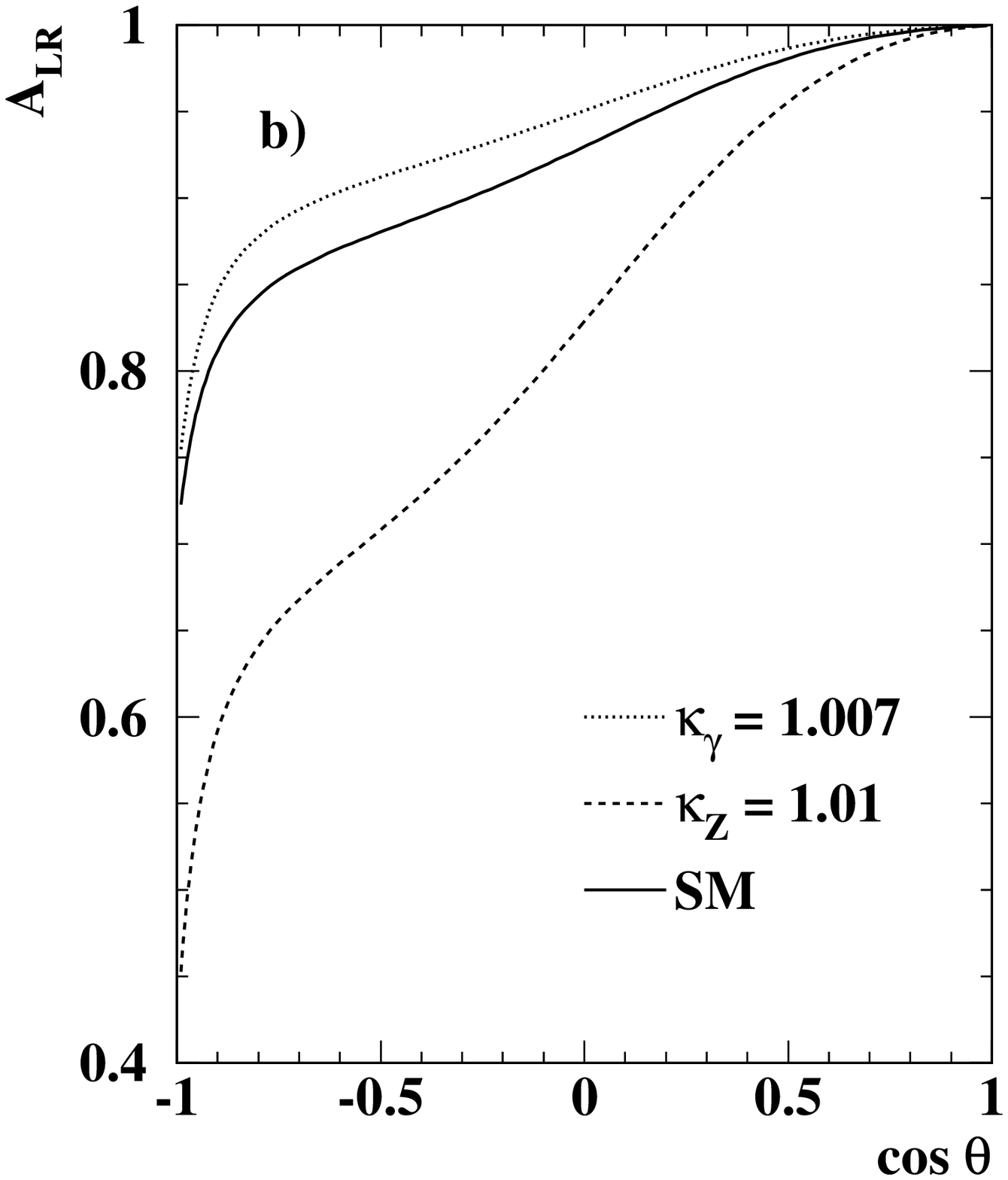}
\end{center}
\caption{Ratio of the differential cross section for W-pair production
  to the Standard Model prediction (a) and left-right asymmetry for
  this process
  (b) as a function of the W-production angle for anomalous 
  $\Ckg$ or $\Ckz$.
  }
\label{fig:tgcpol_pr}
\end{figure}

If the triple gauge couplings are modified by 1-loop corrections one
expects deviations of the order 
$g^2/16 \pi^2 \approx 2.7 \cdot 10^{-3}$.
As an example figure \ref{fig:dkmssm} \cite{ref:kneur} shows the
expected effects on $\kappa_\gamma$ from loop corrections in the MSSM.

\begin{figure}[htbp]
  \begin{center}
    \begin{picture}(10,10)
      \put(0,0){\mbox{\includegraphics[height=10.cm,bb=88 336 504 702]
          {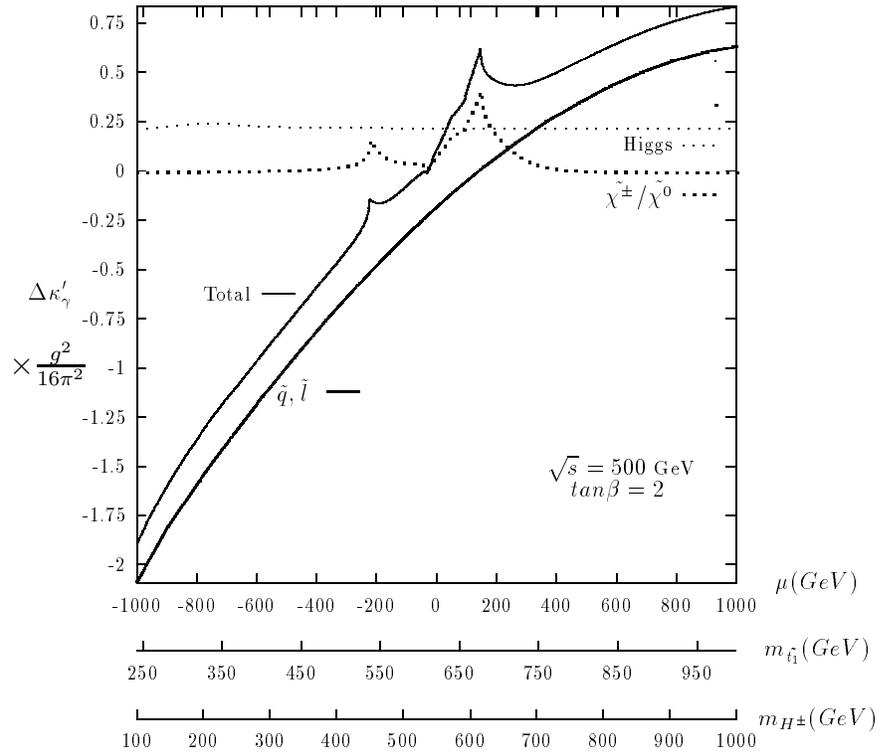}}}
      \put(-0.2,5.){\mbox{{{$\times \frac{g^2}{16 \pi^2}$}}}}
    \end{picture}
  \end{center}
    \caption{Expected effects on $\kappa_\gamma$ from MSSM loop corrections}
    \label{fig:dkmssm}
\end{figure}
\subsection{Experimental procedures}
In $\ee \rightarrow \WW$ anomalous gauge couplings show up in a
modification of the differential cross section with respect to the
scattering angle, where the modification depends on the W polarisation
state. The W-polarisation is accessible from the decay angles of the
W-decay products, so that in total five observables per event are available
(see figure \ref{fig:angdef}):
\begin{itemize}
\item the W production angle $\Theta$;
\item the decay angles $\theta^*$ of the fermions with respect to the W
  flight direction in the W rest frame, sensitive to the longitudinal
  polarisation; 
\item the decay angles $\phi^*$ of the fermions in the W-beam plane,
  sensitive to transverse polarisations.
\end{itemize}
\begin{figure}[htb]
\begin{center}
\includegraphics[height=4cm]{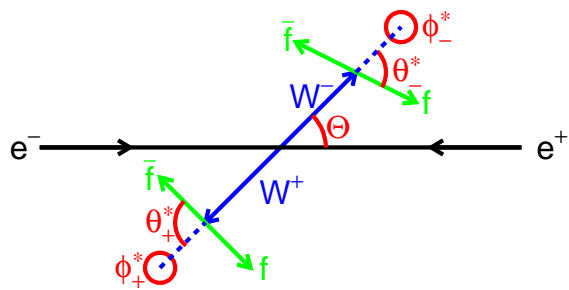}
\end{center}
\caption{Observables available in a W-pair event}
\label{fig:angdef}
\end{figure}

About 46\% of the W-pairs decay into four quarks, 43\% into two quarks
a charged lepton and a neutrino and 11\% into two charged leptons and
two neutrinos. Neglecting initial state radiation and beamstrahlung
the Ws are exactly back to back. In this approximation 
the axis of the W production in the mixed
decays is measured from the decay products
of the hadronically decaying W and the sign ambiguity is resolved by the charge
of the lepton. For the leptonically decaying W both decay angles can
be measured uniquely. For the hadronically decaying W the quark and
the antiquark cannot be distinguished in general, so that the ambiguity
$(\theta^*,\phi^*) \rightarrow (\pi-\theta^*,\phi^*+\pi)$ remains.
If the leptonically decaying W decays into $\tau \nu_\tau$ the resolution
is worse, due to the additional neutrinos, but the measurement is
still possible.
For the fully hadronic W-pairs, even if the jet-pairing is done
correctly, the decay angle ambiguities remain for both Ws and in addition the
$\rm{W}^+$ and the $\rm{W}^-$ cannot be separated, so that these
events are only of very limited use. The ambiguities can be solved
partly by jetcharge techniques or c-tagging with some c-charge
determination, but these methods have only a limited efficiency.
If both W-bosons decay leptonically the full information can be reconstructed
with a twofold ambiguity, if for the lepton-neutrino pairs the W mass
is imposed. This works, however, only if no $\tau$ is involved, which
is in about half of the events.

To analyse the data a binned analysis in a five-dimensional space is
quite impractical. To circumvent this problem several methods are in
use instead \cite{ref:leptwo_tgc}:

\paragraph{Unbinned maximum likelihood fit:} 
In general an unbinned maximum likelihood fit is the statistically most 
powerful method. However it turns out that it is very difficult to
apply corrections for detector effects, backgrounds etc., so that
this fit has not been used for gauge coupling analyses on real data
up to now.

\paragraph{Optimal observable methods:}
If the differential cross section 
$d \sigma(q_1, ... q_m) / d \Omega\ 
(\Omega=(\theta_1, ... \theta_n))$ is linear in the 
$m$ parameters, $q_j$, to be determined,
i.e. the differential cross section can be written as 
\[
\frac{d \sigma}{d \Omega} = S_0(\Omega)  + \sum_j S_{1,j}(\Omega)  q_j
\]
one can construct an optimal observable 
${\cal O}_j(\Omega) = S_{1,j}(\Omega)  / S_0(\Omega)  $
for which it can
be shown that the $q_j$ can be extracted in a statistically optimal way
from the $\langle {\cal O}_j \rangle$ \cite{ref:opt1,ref:opt2}.
If instead the distributions of the ${\cal O}_j(\Omega)$ are fitted,
effects due to non linearity are accounted for automatically
and for detector effects one can correct in the usual way.
The ${\cal O}_j(\Omega)$ is just no longer exactly optimal, but when 
the non-linearity and detector corrections are small 
this is usually negligible.
For $m<n$ one can thus reduce the
number of parameters without a loss in statistical precision.

\paragraph{Spin density matrix:}
The full information of the W pair cross section can be expressed as
the product of the differential cross section 
$\partial  \sigma/\partial \cos \Theta$
and the spin density matrix
\begin{equation}
 \rho_{\tau_-\tau'_- \tau_+\tau'_+ }(\cos \Theta) = 
   \frac{  
     \mathcal{F}_{\tau_-\tau_+}^{(\lambda)}
 \left( \mathcal{F}_{\tau'_-\tau'_+}^{(\lambda)}\right)^{\ast}}
{ \sum_{\tau_+ \tau_-} \left|  \mathcal{F}_{\tau_-\tau_+}^{(\lambda)}\right|^2},
\end{equation}
where $\mathcal{F}_{\tau_-\tau_+}^{(\lambda)}$ is the amplitude
for $\rm W^-\rm W^+$-pairs with W polarisations $\tau_-$ and $\tau_+$,
respectively, and an initial electron polarisations of 
$\lambda$\footnote{
For unpolarised beams one has to sum over $\lambda$
}.
In total this system has 81 density matrix elements where one is given by
the  normalisation 
\[
 \sum_{\tau_+ \tau_-}  \rho_{\tau_-\tau_- \tau_+\tau_+} = 1.
\]
This number reduces to
35 elements in a CP conserving theory, as is described in \cite{BILENKY}.
Only little information is lost if the spin density matrix for single
Ws is used in the analysis, neglecting spin correlations:
\[ 
\rho_{\tau_-\tau_-'}(\cos \Theta) =  
      \sum_{\tau_+} \rho_{\tau_-\tau'_- \tau_+\tau_+ }.
\]
$\rho_{\tau \tau'}$ is hermitian, thus having 6 independent matrix elements.
The diagonal elements, which are real, can be interpreted as the probability to
find a W with helicity $\tau$. The imaginary parts of the off-diagonal
elements vanish if CP is conserved.

Experimentally $\rho_{\tau \tau'}$ can be obtained from 
\[
\rho_{\tau \tau'} \frac{\partial \sigma}{\partial \cos \Theta} =
\int \frac{\partial^3  \sigma}{\partial \cos \Theta \partial \cos \theta^* 
\partial \phi^*}
\Lambda_{\tau \tau'}(\theta^*,\phi^*) d \cos \theta^* d \phi^*
\]
The $\Lambda_{\tau \tau'}$ are projection operators for the different
helicity states, for example 
$\Lambda_{00}=2 - 5 \cos^2 \theta^*$.
In this way the initial 5-dimensional parameter space can be reduced to 6
one-dimensional distributions.
\subsection{Results at LEP and the {\sc Tevatron}}
Triple gauge couplings have already been measured at LEP2
and at the {\sc Tevatron}.
At LEP2 the measurements use the same processes as a linear collider
will use.
Since there one is closer to threshold the scales for W-pair and
single W production are not so much different. However the single W
production cross section is much smaller, so that also at LEP the
dominant information comes from pair production.
In $\pp$-collisions the most sensitive process is W$\gamma$-pair
production, where the high $p_t$ photon is radiated off a W. Also WZ-
and WW-pairs are analysed, however with a lower sensitivity than
the W$\gamma$-pairs. $\pp$-collisions are thus mainly sensitive to
WW$\gamma$ couplings.
In general all measurements agree well with the Standard Model
prediction with a precision of a few percent, so that in no
realistic model a deviation is expected.
As an example figure \ref{fig:leptgc} shows the results of the three
parameter fit in the  $(\Delta g_1^Z,-\Delta \kappa_\gamma)$ and
$(\Delta g_1^Z-\lambda_\gamma)$ plane\cite{ref:ewppe}.
\begin{figure}[htbp]
\begin{center}
  \includegraphics[width=\textwidth]{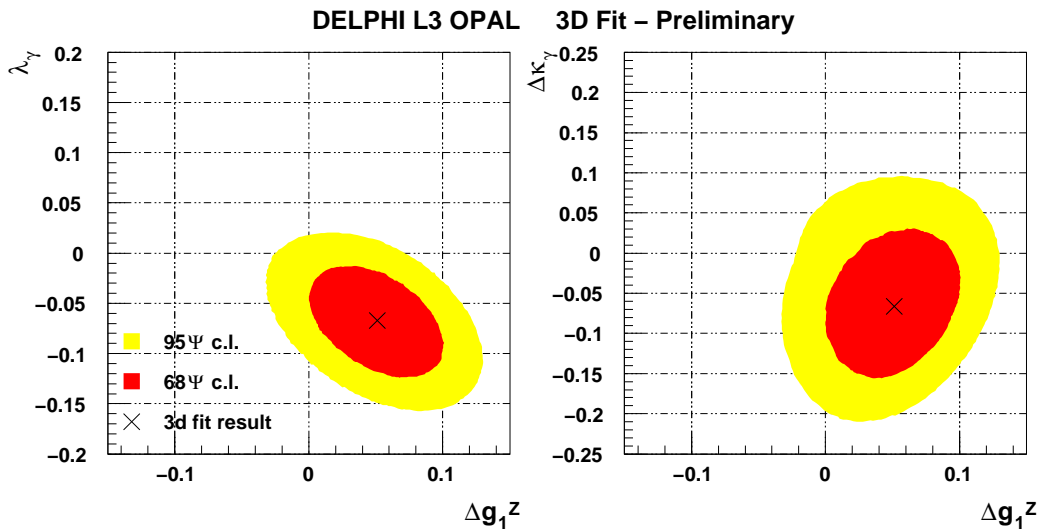}
\end{center}
\caption{Result of the 3-parameter fit to the triple gauge couplings at 
DELPHI, L3 and OPAL.
}
\label{fig:leptgc} 
\end{figure}

\subsection{Expectation from the linear collider}
For the linear collider studies the spin density matrix formalism has
been used which produces close to optimal results \cite{ref:menges}. 
Fits have been
performed for several numbers of free parameters with and without beam
polarisation. As already said the symmetries of the couplings manifest
themselves in the spin density matrix elements that are non-zero, so
that the sets of couplings with different symmetries are basically
measured without correlations between the sets. 
Table \ref{tab:singtgc} shows the results of the different single parameter
fits including the C or P violating couplings
for $500 \fbi$ at $\sqrt{s}=500 \GeV$ and $1000 \fbi$ at $\sqrt{s}=800 \GeV$.
For both cases an electron polarisation of $\pm 80\%$ and a positron 
polarisation
of $\pm 60\%$ is assumed.
Figure \ref{fig:fiveparfit} shows the results of the five-parameter fit for
$\sqrt{s} = 800 \GeV$. Only the combinations with large correlations are shown.

\begin{table}
\begin{center}
\begin{tabular}[c]{|c|c|c|}
\hline
coupling & $\sqrt{s}=500\GeV$ & $\sqrt{s}=800\GeV$ \\
\hline
\multicolumn{3}{|l|}{C,P-conserving, $SU(2)\times U(1)$ relations:}\\
\hline
  \Cdgz  &$  2.8 $&$  1.8 $\\
  \Cdkg  &$  3.1 $&$  1.9 $\\
  \Clg   &$  4.3 $&$  2.6 $\\
\hline
\multicolumn{3}{|l|}{C,P-conserving, no relations:}\\
\hline
  \Cdgz  &$ 15.5 \phantom{0} $&$ 12.6 \phantom{0} $\\
  \Cdkg  &$  3.3 $&$  1.9 $\\
  \Clg   &$  5.9 $&$  3.3 $\\
  \Cdkz  &$  3.2 $&$  1.9 $\\
  \Clz   &$  6.7 $&$  3.0 $\\
\hline
\multicolumn{3}{|l|}{not C or P conserving:}\\
\hline         
  \Cgz{5}&$ 16.5 \phantom{0} $&$ 14.4 \phantom{0} $\\
  \Cgz{4}&$ 45.9 \phantom{0} $&$ 18.3 \phantom{0} $\\
  \Ckzt  &$ 39.0 \phantom{0} $&$ 14.3 \phantom{0} $\\
  \Clzt  &$  7.5 $&$  3.0 $\\
  \hline
\end{tabular}
\end{center}
\caption{Expected uncertainties of the single parameter fits to the different 
triple gauge 
couplings. For $\sqrt{s}=500 \GeV\ {\cal L}=500\fbi$ and for
$\sqrt{s}=800 \GeV\ {\cal L}=1000\fbi$ has been assumed. For both energies
$\pmi = 80\%$ and $\ppl = 60\%$ has been used.
}
\label{tab:singtgc} 
\end{table}
\begin{figure}[htbp]
\begin{center}
\includegraphics[height=6.cm,bb=2 8 542 528]{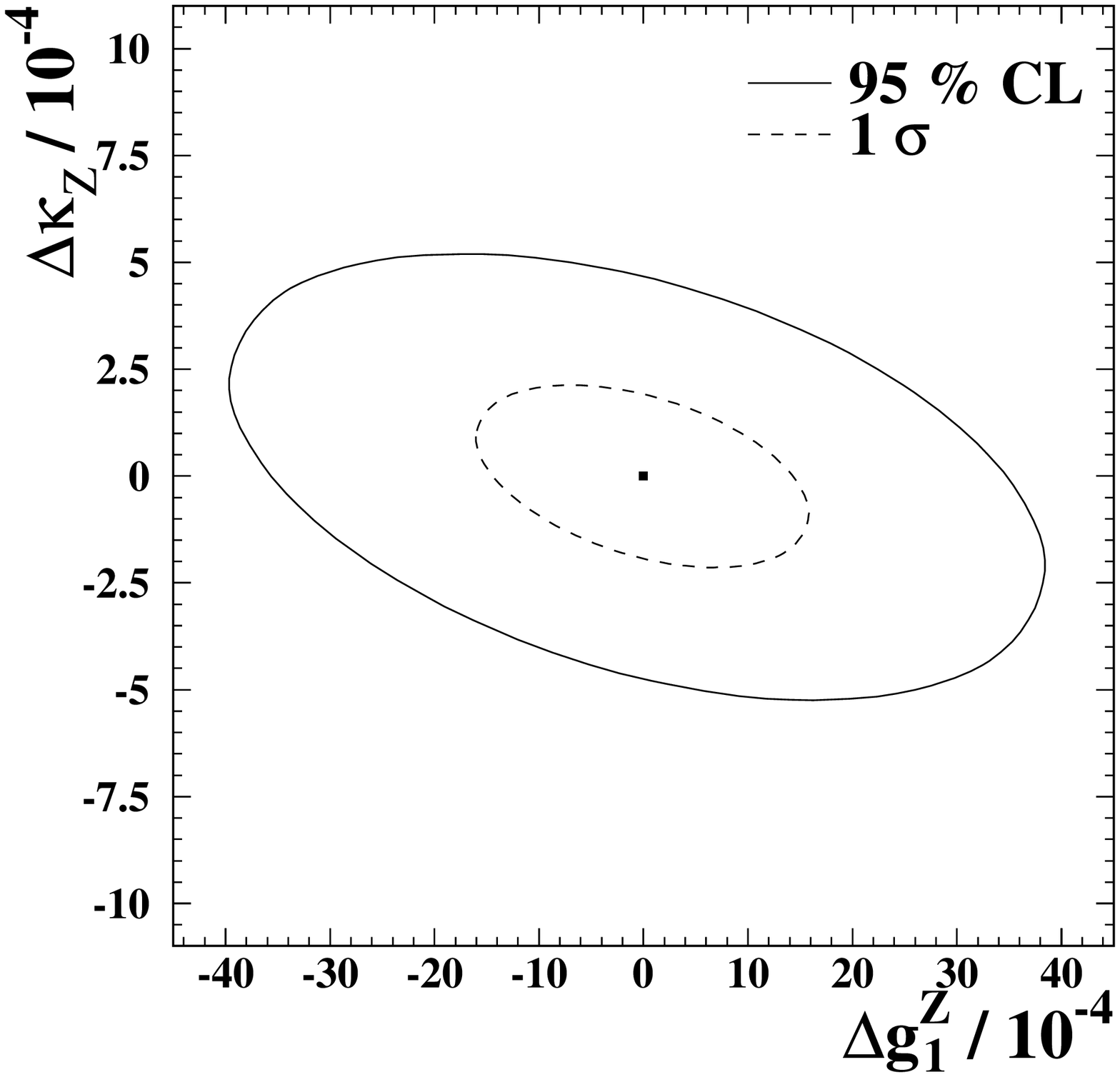}
\includegraphics[height=6.cm,bb=2 8 542 528]{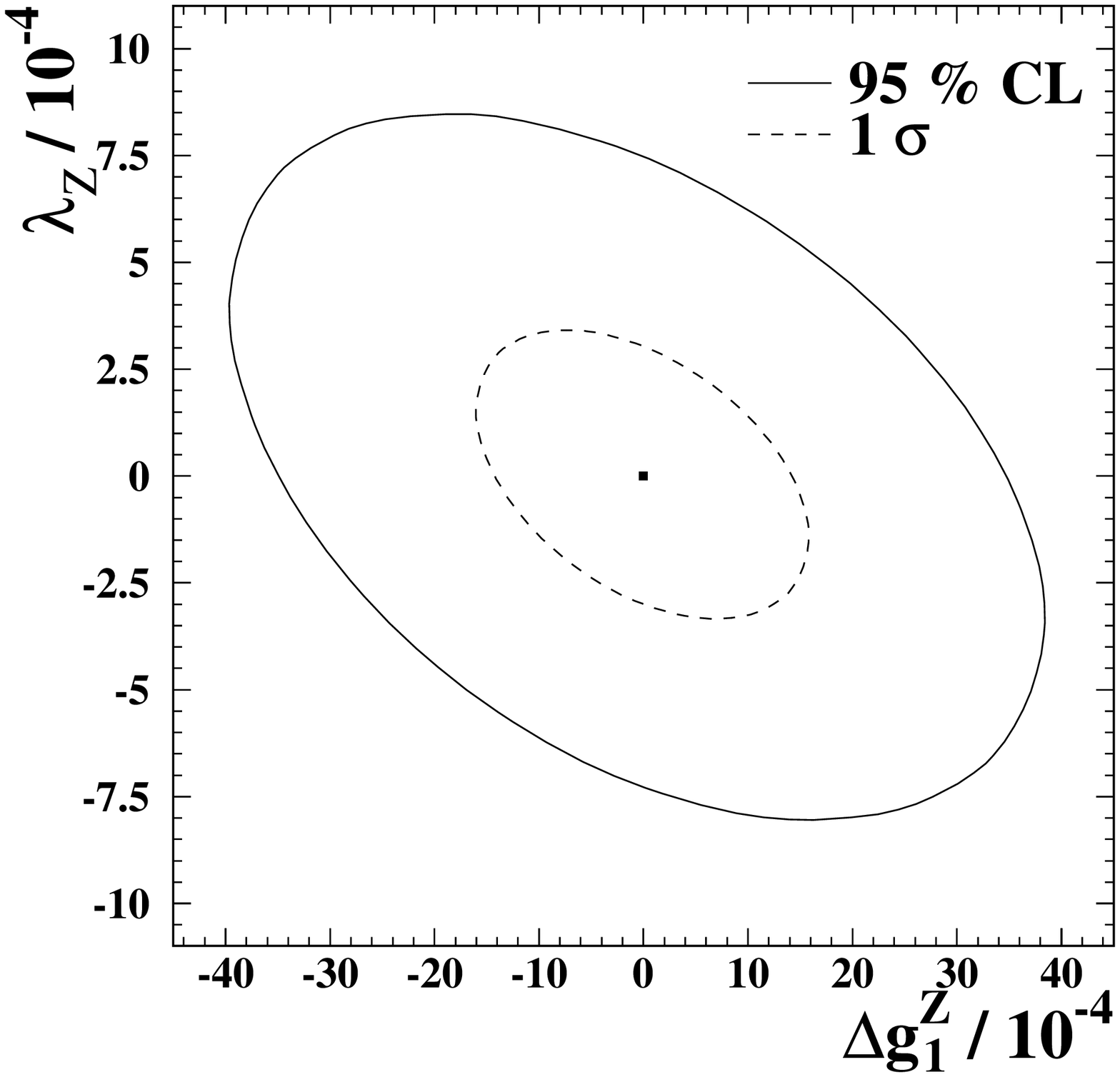}
\includegraphics[height=6.cm,bb=2 8 542 528]{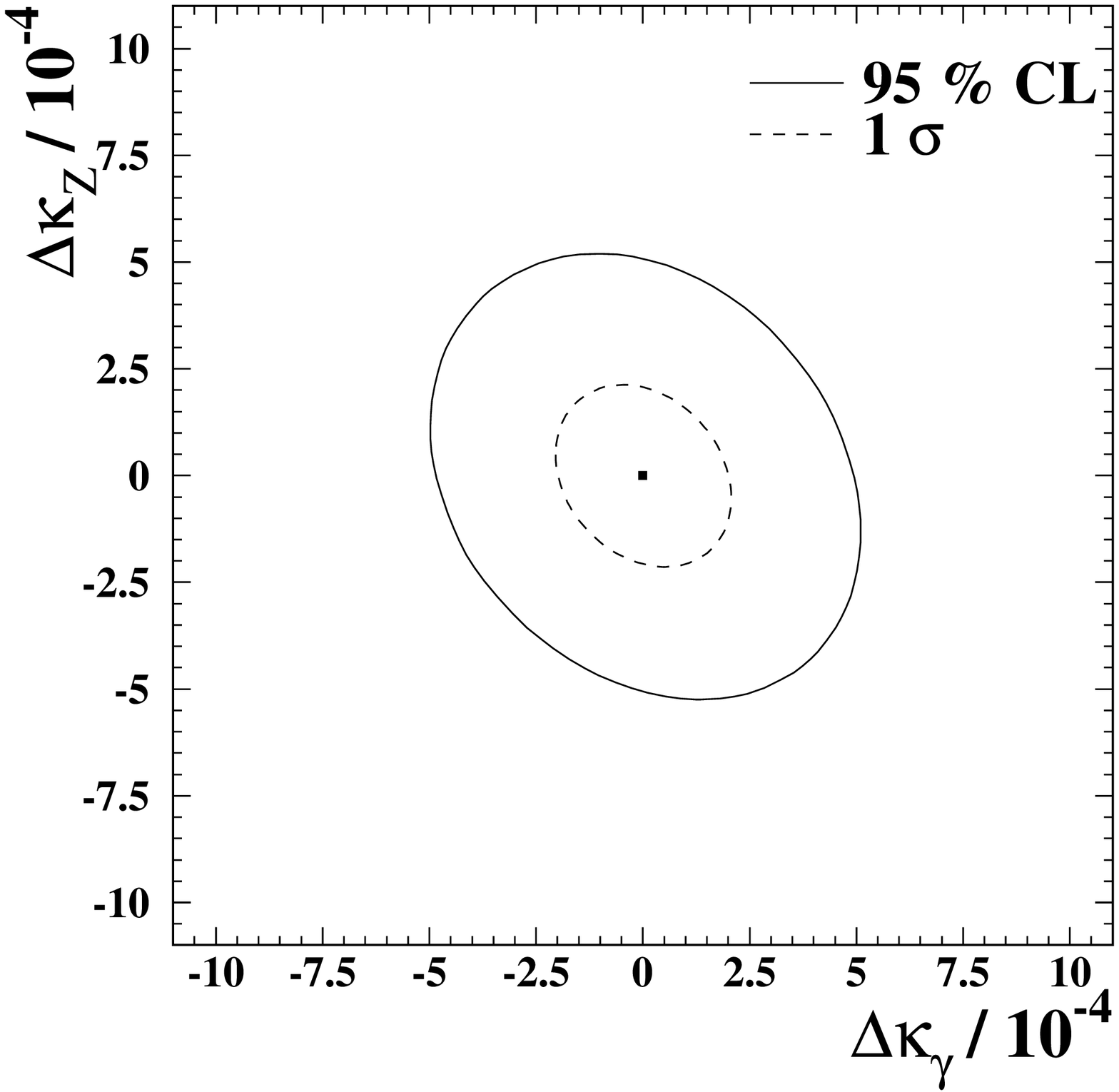}
\includegraphics[height=6.cm,bb=2 8 542 528]{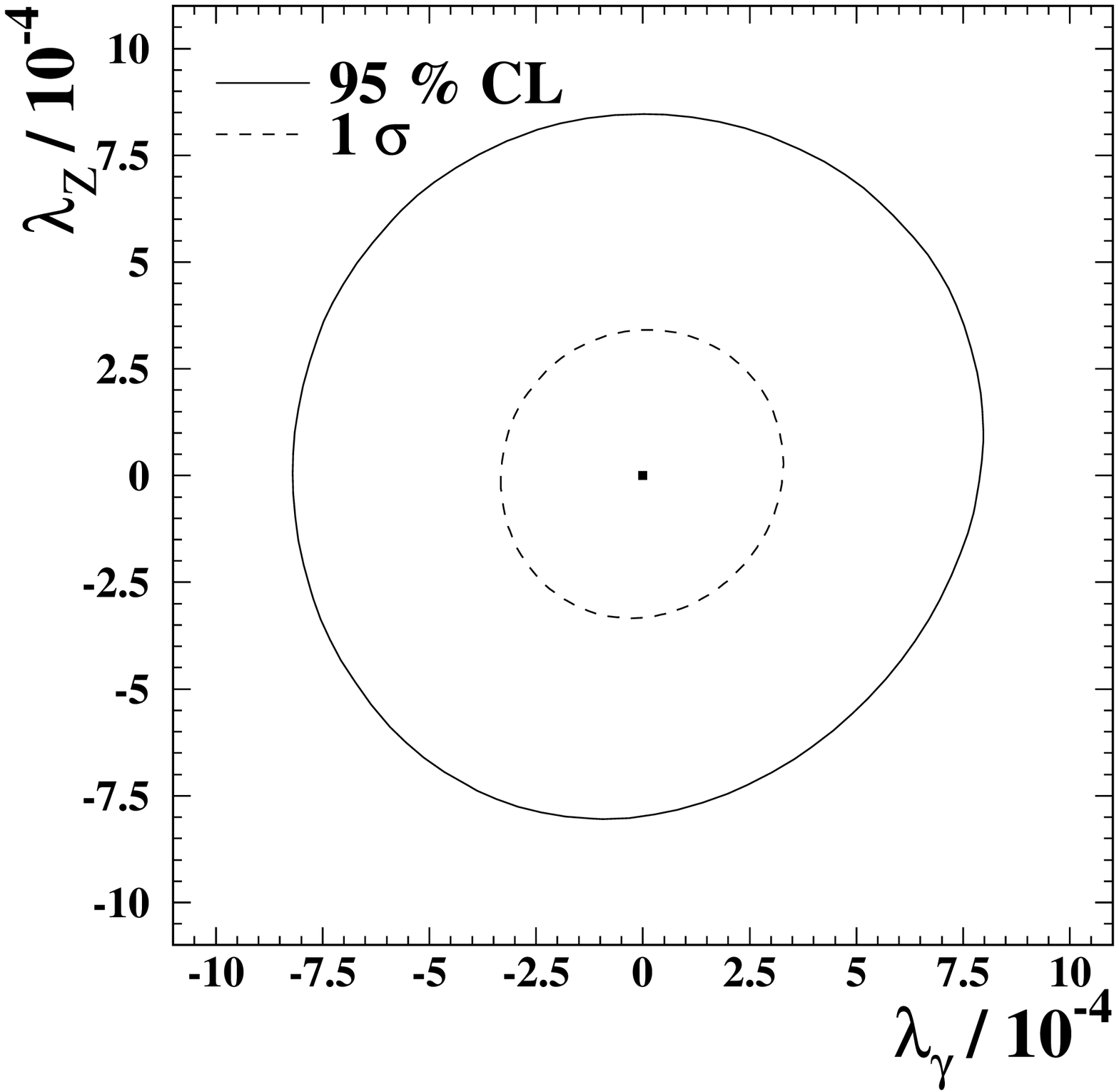}
\end{center}
\caption{$1 \sigma$ and $95\%$ c.l. (2D) contours for \Cdgz-\Cdkz,
  \Cdgz-\Clz, \Cdkg-\Cdkz and \Clg-\Clz in the 5-parameter fit 
  ($\sqrt{s}=800 \GeV, \, {\cal L} = 1000 \fbi,\, \pmi=0.8,\,\ppl=0.6$).
  For the combinations not shown the correlations are small.
}
\label{fig:fiveparfit} 
\end{figure}

Due to the large boost the resolution of the W-production angle is
much better than at LEP. 
For the same reason the resolution in the
decay angles is somewhat worse, however only sin and cos functions
have to be distinguished here, so that the resolution is largely
sufficient. In total the corrections due to detector effects are
almost negligible and all experimental systematics are small.
The radiative corrections however need to be known to significantly
better than 1\%. The beam polarisations can be measured from an extension 
of the Blondel scheme. The large forward peak of the cross section is
dominated by neutrino t-channel exchange, so that its left-right
asymmetry is very close to one, independent of the triple
couplings. This region can be used to measure the electron
polarisation directly from data, even if no positron polarisation is available.

Without beam polarisation only the fits without separating WW$\gamma$
and WWZ couplings are possible. For these fits the errors increase by
about a factor of two.

The precision of $\kappa_\gamma$ for $\sqrt{s} = 500 \GeV$ for example
is $0.12 \cdot (g^2 /16 \pi^2)$. If this is compared to the expected loop
corrections from the MSSM, 
shown in figure \ref{fig:dkmssm} one sees,
that the gauge couplings can probe extensions of the Standard Model
with sensitivities similar to the Z-fermion couplings.

Figure \ref{fig:tgccomp} compares the expected errors on 
$\kappa_\gamma$ and $\lambda_\gamma$ at the linear collider 
with those from other machines.
Especially for $\kappa_\gamma$ the precision is much better than at
the LHC. 
Since the dimension of the corresponding operator is lower for
$\kappa$ than for $\lambda$ the suppression due to the high scale of
new physics should be smaller. This parameter should thus be more
sensitive to new effects for example from strong electroweak symmetry breaking.

\begin{figure}[htb]
\begin{center}
\includegraphics[width=0.48\linewidth,bb=10 68 492 485]{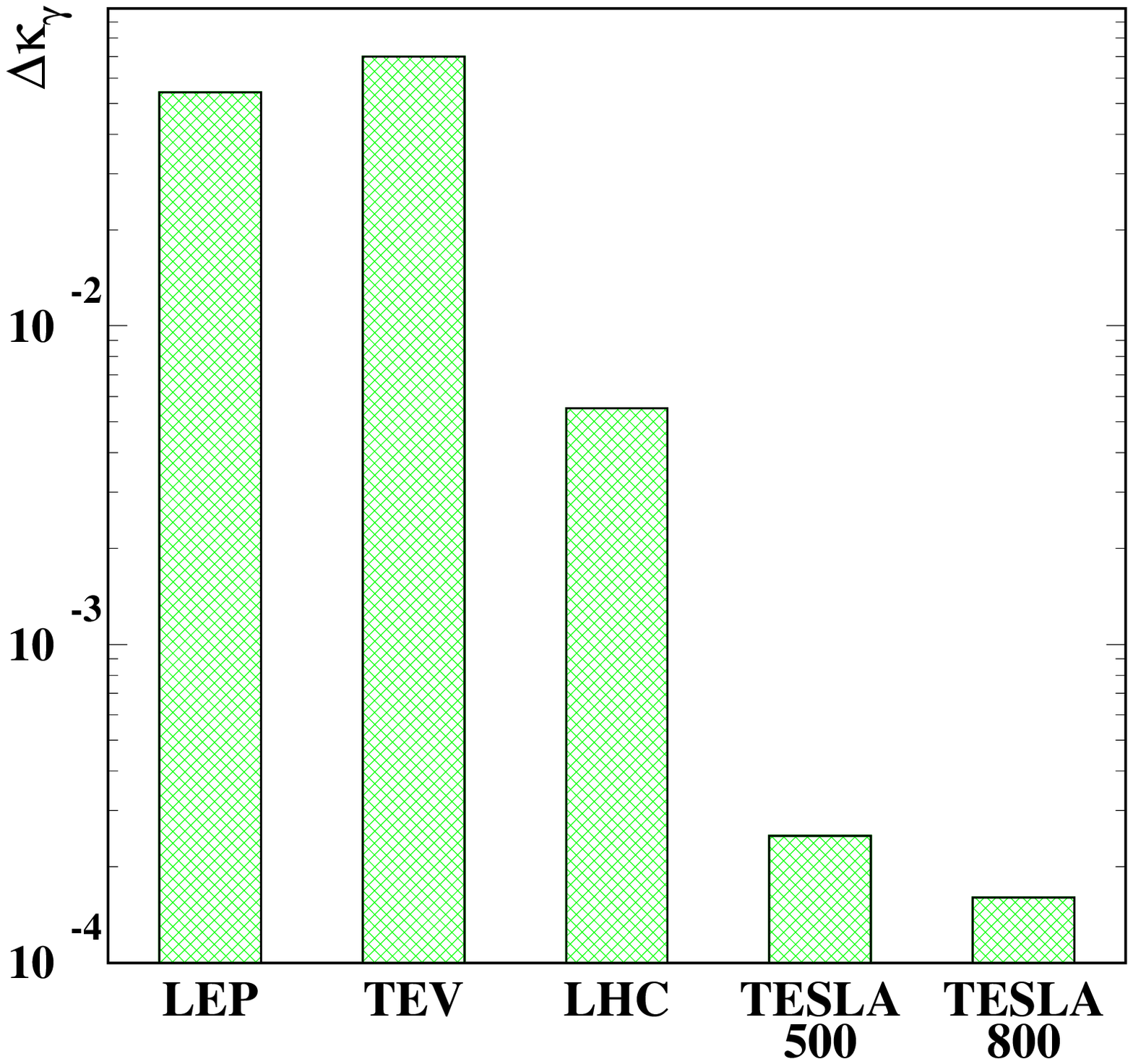}
\includegraphics[width=0.48\linewidth,bb=10 68 492 485]{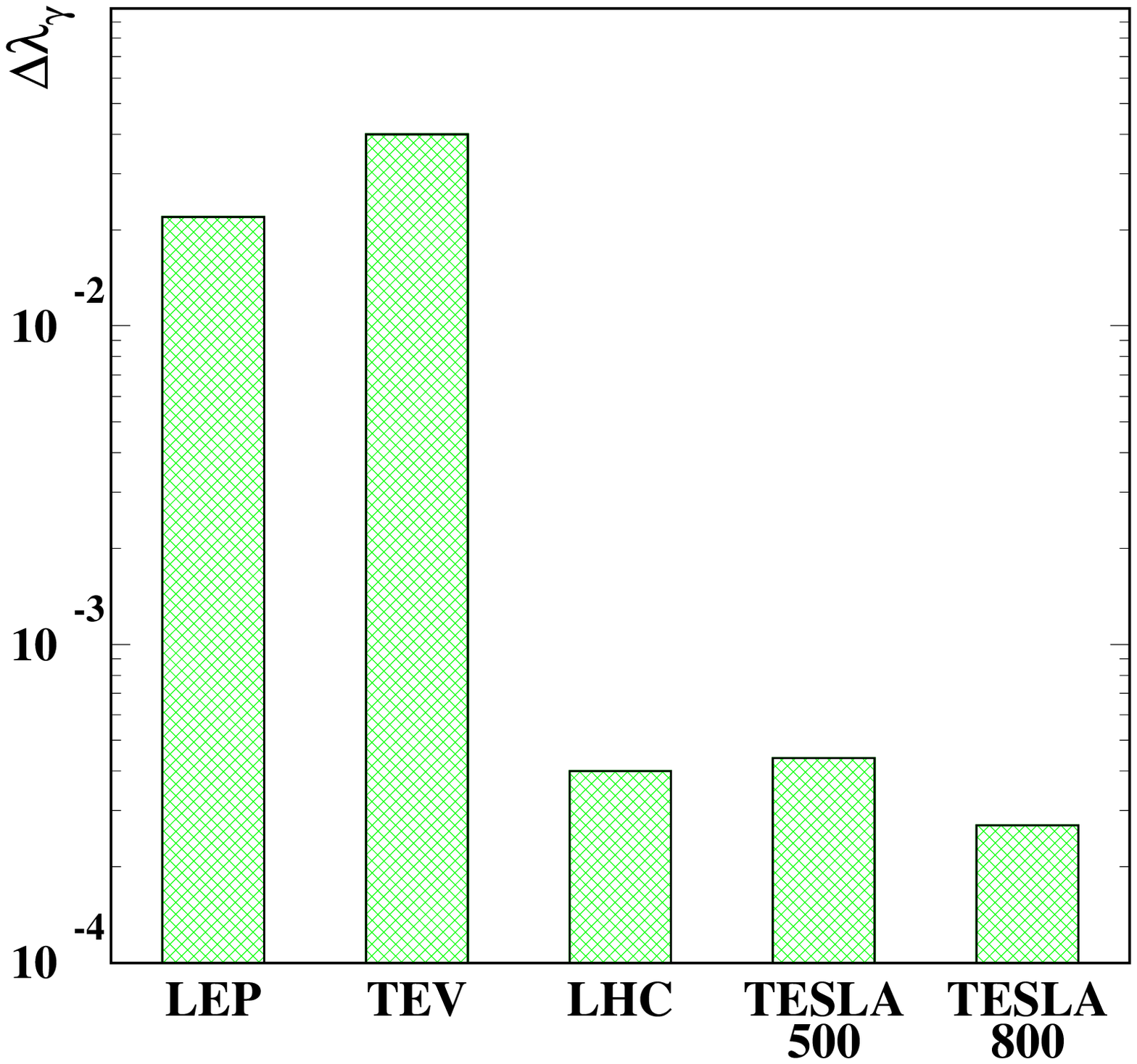}
\end{center}
\caption{Comparison of $\Delta \kappa_\gamma$ and $\Delta
  \lambda_\gamma$ at different machines. For LHC and TESLA three years
  of running are assumed (LHC: $300\fbi$, 
  TESLA $\sqrt{s}=500\GeV$: $900 \fbi$, TESLA $\sqrt{s}=800\GeV$: $1500 \fbi$).
  }
\label{fig:tgccomp}
\end{figure}

Also the possibility to search for the neutral triple gauge couplings
Z$\gamma \gamma$ and ZZ$\gamma$ has been studied \cite{juan}. However
the possible limits are about an order of magnitude worse than
predictions from loop effects in the Standard Model and expectations
from the LHC. For this reason they will not be discussed further.

\section{Strong electroweak symmetry breaking}
If no Higgs boson exists electroweak interactions become strong at
high energy and WW scattering violates unitarity at 
$\sqrt{s_{\rm{WW}}} = 1.2 \TeV$. The latest at this energy new
physics has thus to set in, but with precision measurements it should be
visible already at lower energies.

Strong electroweak symmetry breaking has already been discussed in
detail in chapter 5\cite{ref:wolfgang}, 
so only the features relevant to gauge boson
physics will be repeated here.
Without a Higgs the longitudinal degrees of freedom of the W
and the Z become the Goldstone bosons 
of a strongly interacting theory
and the low energy equivalence theorem (LET) \cite{ref:let1,ref:let2}
relates longitudinal
vector boson scattering to pion scattering at low energy.
Concrete models typically predict resonances 
like the $\rho$ or the
$\omega$ in QCD. However, as shown in figure \ref{fig:lepint} an exact
copy of QCD is already excluded by the LEP/SLD
precision data.

In a similar way as the $\rho$-resonance shows up in 
$\ee \rightarrow \pi^+ \pi^-$ also W-pair production should be
sensitive to the strongly interacting scenario. In a model dependent
analysis the amplitude for the production of two longitudinal Ws has
been multiplied by a QCD like form factor assuming a resonance mass
and width \cite{timtrho}. 
The LET is then the asymptotic value for high resonance masses.
Figure \ref{fig:techni} shows the expected sensitivity with
$\sqrt{s} = 800 \GeV$ and ${\cal L} = 500 \fbi$. Within the assumed
model the LET and the SM model can clearly be distinguished and one is
sensitive to resonance masses up to around $3 \TeV$. 
\begin{figure}[htbp]
\begin{center}
\begin{turn}{-90}
\includegraphics[height=8.5cm]{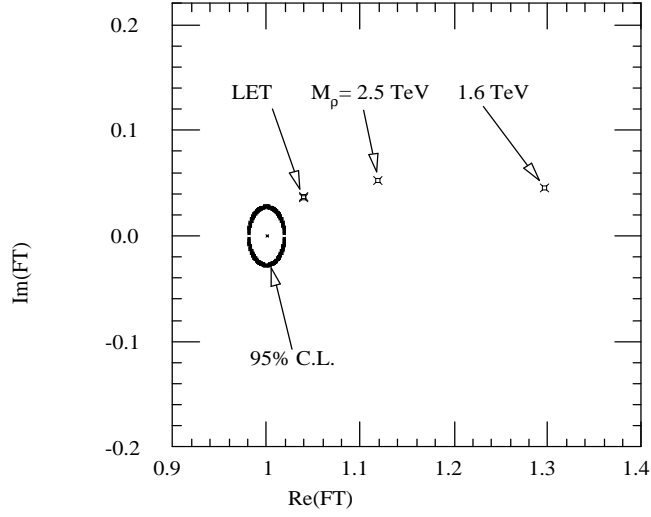}
\end{turn}
\end{center}
\caption{
Sensitivity to a $\rho$ like resonance in W-pair production at
$\sqrt{s} = 800 \GeV$ and ${\cal L} = 500 \, \fbi$.
}
\label{fig:techni} 
\end{figure}

To analyse the data in a model independent way effects from strong
electroweak symmetry breaking can be parameterised in an effective Lagrangian.
From LEP1/SLD it is known that the $\rho$-parameter 
is approximately
one and the small deviations can fully be described by Standard Model
loop corrections, mainly from the top quark. It is therefore reasonable
to keep only terms that conserve the custodial $SU(2)_c$ 
symmetry.
Under this assumption the effective Lagrangian for the
triple gauge couplings, 
keeping only terms of the lowest dimension, reads: 
\begin{eqnarray*}
L_{TGC} & = & \frac{\alpha_{1}}{16\pi^2} \frac{gg'}{2} B_{\mu\nu} 
                \mathop{\mathrm{tr}} \Bigl( \sigma_3 W^{\mu\nu} \Bigr) +\\
        &   & \frac{\alpha_{2}}{16\pi^2} \mathrm{i}g' B_{\mu\nu}
                \mathop{\mathrm{tr}} \Bigl( \sigma_3 V^\mu V^\nu \Bigr) +\\
        &   & \frac{\alpha_{3}}{16\pi^2} 2\mathrm{i}g
                \mathop{\mathrm{tr}} \Bigl( W_{\mu\nu} V^\mu V^\nu \Bigr) 
\end{eqnarray*}
with $V_\mu = -\mathrm{i}g\frac{\sigma^i}{2}W^i_\mu
+\mathrm{i}g'\frac{\sigma^3}{2}B_\mu$.
From a naive dimensional analysis one expects 
\[
  \frac{\alpha_i}{16\pi^2} = \left(\frac{v}{\Lambda^*_i}\right)^2
\]
with $v=246 \GeV$ and $\Lambda^*_i$ being the scale of the new physics.
From unitarity arguments one needs $\Lambda^* \sim 3 \TeV$ so that 
the $\alpha_i$ should be of ${\cal O}(1)$.
The $\alpha_i$ can be expressed in terms of $g_1^Z,\, \kappa_\gamma, \kappa_Z$
as:
\begin{eqnarray}
  g_1^Z &=& 1 + \frac{e^2}{\cos^2\theta_W(\cos^2\theta_W\!-\!\sin^2\theta_W)}
               \frac{\alpha_{1}}{16\pi^2}
             + \frac{e^2}{ \sin^2\theta_W\cos^2\theta_W}
               \frac{\alpha_{3}}{16\pi^2} \nonumber \\
  \kappa_\gamma &=& 1
                   - \frac{e^2}{\sin^2\theta_W}\frac{\alpha_{1}}{16\pi^2}
                   + \frac{e^2}{\sin^2\theta_W}\frac{\alpha_{2}}{16\pi^2}
                   + \frac{e^2}{\sin^2\theta_W}\frac{\alpha_{3}}{16\pi^2} 
               \label{eq:kaptoalph}\\ 
  \kappa_Z &=& 1 + \frac{2e^2}{\cos^2\theta_W\!-\!\sin^2\theta_W}
                  \frac{\alpha_{1}}{16\pi^2}
                - \frac{e^2}{\cos^2\theta_W}\frac{\alpha_{2}}{16\pi^2}
                +
               \frac{e^2}{\sin^2\theta_W}\frac{\alpha_{3}}{16\pi^2}
               \nonumber
\end{eqnarray}
Unfortunately eq. \ref{eq:kaptoalph} is singular in the blind direction
\[
  (\alpha_{1},\; \alpha_{2},\; \alpha_{3})_{{\rm blind}}
    \propto
  (\cos^2\theta_W-\sin^2\theta_W,\; \cos^2\theta_W,\; -\sin^2\theta_W)
\]
so that not all the $\alpha_i$ can be extracted from triple gauge couplings.
However, $\alpha_1$ can also be constrained from the Z-pole data. 
Figure \ref{fig:alpha_123} shows the expected errors in
the $(\alpha_1,\alpha_2,\alpha_3)$ space for $\sqrt{s}=800\GeV,\,
{\cal L}=100 \, \fbi, \pmi = 80\%$ and $\ppl = 60\%$ and from the Z-pole
expectation at Giga-Z.
Also the errors in $(\alpha_2,\alpha_3)$ for $\alpha_1 = 0$ and for the
combination of the high- and low-energy data are shown. The
errors translate into mass scale limits of 
$\Lambda^*_i \approx 10 \TeV$ which are well above the unitarity limit
of $3 \TeV$.

\begin{figure}[htbp]
\begin{center}
\includegraphics[width=\linewidth]{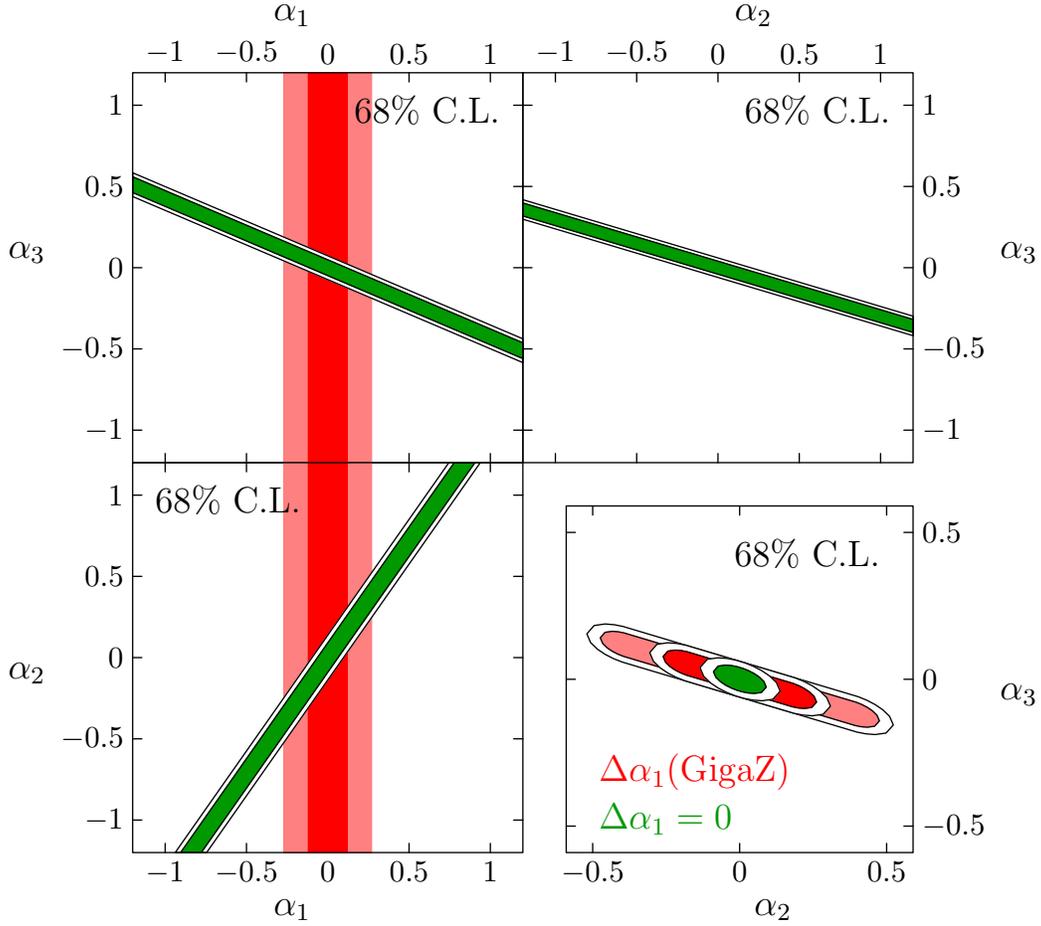}
\end{center}
\caption{
Sensitivity to the effective couplings $\alpha_{1,2,3}$ from the
measurement of the triple gauge couplings at a linear collider and from
the Z-pole precision data. 
For the Z-pole constraint the outer region is without and the inner region 
with the accurate $\MW$ measurement.
In the lower right plot the $\alpha_2-\alpha_3$ plane with $\alpha_1=0$ 
or with the Z-pole constraint is shown.
}
\label{fig:alpha_123} 
\end{figure}

In the description of the quartic couplings 
with an effective
Lagrangian there are two operators that conserve $SU(2)_c$
and that do not contribute already to the triple couplings:
\begin{eqnarray*}
L_{QGC} & = & \frac{\alpha_{4}}{16\pi^2}
              \mathop{\mathrm{tr}} \Bigl( V_\mu V_\nu \Bigr)
              \mathop{\mathrm{tr}} \Bigl( V^\mu V^\nu \Bigr) +\\
        &   & \frac{\alpha_{5}}{16\pi^2}
              \mathop{\mathrm{tr}} \Bigl( V_\mu V^\mu \Bigr)
              \mathop{\mathrm{tr}} \Bigl( V_\nu V^\nu \Bigr)
\end{eqnarray*}
Experimentally the quartic couplings are measured through processes as
sketched im figure \ref{fig:vvscatfeyn}. Since the electron coupling to the 
W is about a factor of two larger than the one to the Z, 
the sensitive processes
have two neutrinos in the final state. This means that the gauge bosons in the
final state have to be fully reconstructed so that decay modes involving
neutrinos cannot be used.



In principle there are three processes sensitive to $\alpha_4$ and $\alpha_5$:
\begin{eqnarray*}
  \ee & \rightarrow & \nu \nu \WW\\
  \ee & \rightarrow & \nu \nu \rm{ZZ}\\
  \rm{e}^- \rm{e}^-& \rightarrow & \nu \nu \rm{W}^-\rm{W}^- 
\end{eqnarray*}
The sensitivity for the three processes is similar with different
dependence on the two couplings. However, since the luminosity for 
$\rm{e}^- \rm{e}^-$ is about an order of magnitude less than for 
$\ee$ the sensitivity per unit of running time in this channel is
significantly smaller. A large potential background to these processes is
$  \ee \rightarrow \ee \WW$ proceeding via intermediate photons.
To reject this process it is important that electrons can be vetoed in
the detector down to the lowest possible angles. If the electrons escape in the
beampipe the total transverse momentum of the two-W system has to be very
small. The $  \ee \rightarrow \ee \WW$ events can then be rejected by a 
cut on the total transverse momentum.
Unfortunately such a cut suppresses also
the longitudinal W-pairs more than the transverse ones.
Another important background process is $\ee \rightarrow {\rm e} \nu {\rm WZ}$.
This background is in principle smaller than $\ee \rightarrow \ee \WW$, but
because of the missing neutrino the transverse momentum cut doesn't work.
The process can only be separated from the signal if the energy flow resolution
of the detector for hadronic jets is good enough to separate Ws and Zs.
Another possibility to enhance the signal to background ratio is beam 
polarisation. 
Ws couple only to left-handed electrons and right-handed 
positrons while the electron-photon coupling is polarisation independent.

In a possible analysis signal events are first separated from the background 
and then analysed in terms of the quartic couplings. 
Variables sensitive to these couplings are:
\begin{itemize}
\item the VV (V=W,Z) centre of mass energy;
\item the V production angle in the VV rest frame, to select hard VV-scattering;
\item the V decay angles to select longitudinal Vs.
\end{itemize}
Figure \ref{fig:alpha45} shows the sensitivity of the linear collider
at $\sqrt{s}=800\GeV$ for  $\ee \rightarrow \nu \nu \WW$ and
$\ee \rightarrow \nu \nu \rm{ZZ}$ \cite{ref:roberto}. 
Due to the different dependence on
the two parameters the combination of the two processes greatly
reduces the error. The reach in $\Lambda^*$ is $\Lambda_4^*\approx 1.8 \TeV$
and $\Lambda_5^*\approx 2.5 \TeV$ for the two parameter fit increasing
to $\Lambda_4^*\approx 2.9 \TeV$ and $\Lambda_5^*\approx 4.9 \TeV$ 
for one-parameter fits, just
reaching the interesting region of $3 \TeV$. The sensitivity increases
strongly with the centre of mass energy. A factor of two in energy
reduces the errors on $\alpha_{4,5}$ by almost an order of magnitude.
\begin{figure}[htbp]
\begin{center}
  \includegraphics[width=0.45\textwidth,height=0.45\textwidth]{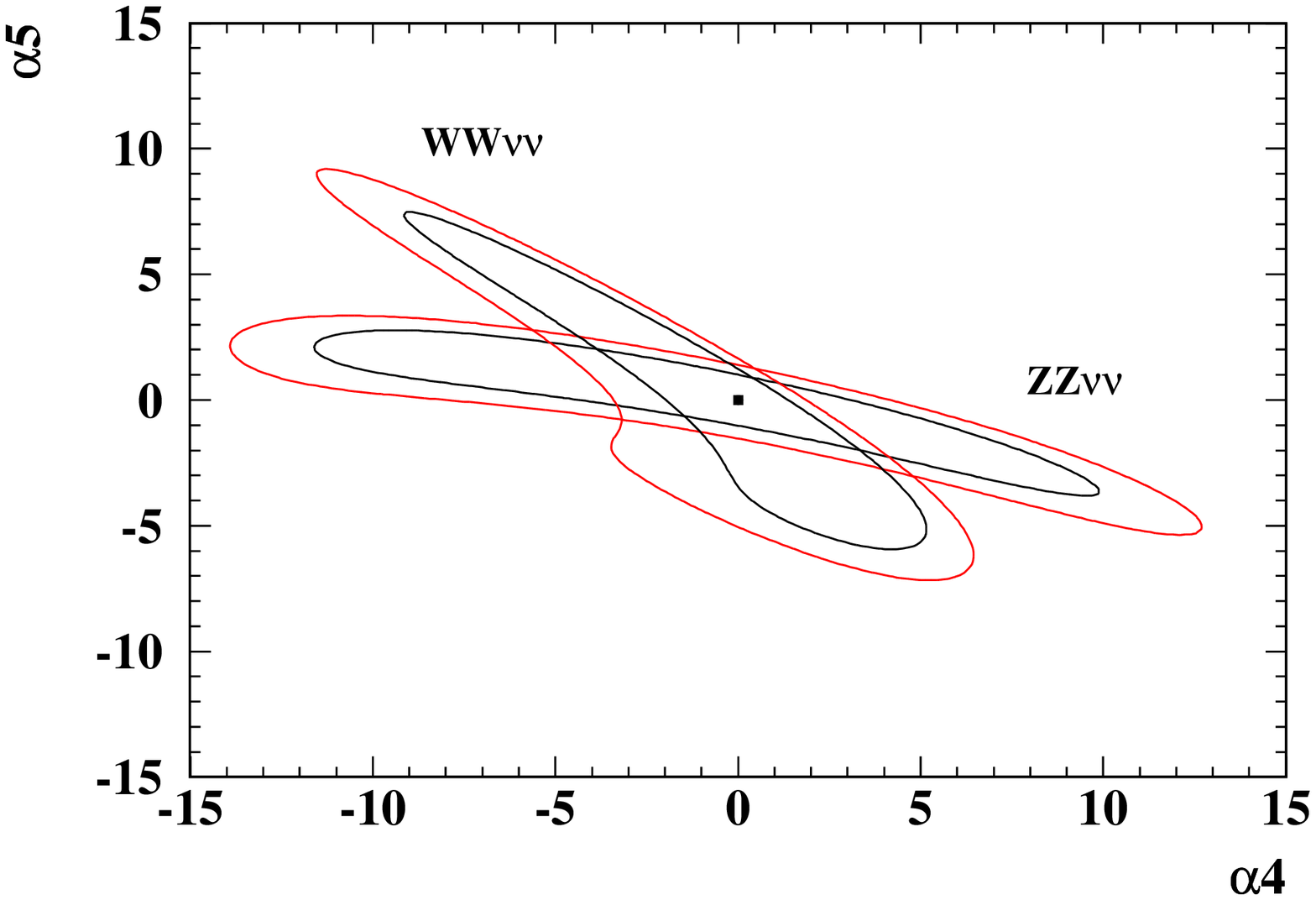}\qquad
  \includegraphics[width=0.45\textwidth,height=0.45\textwidth]{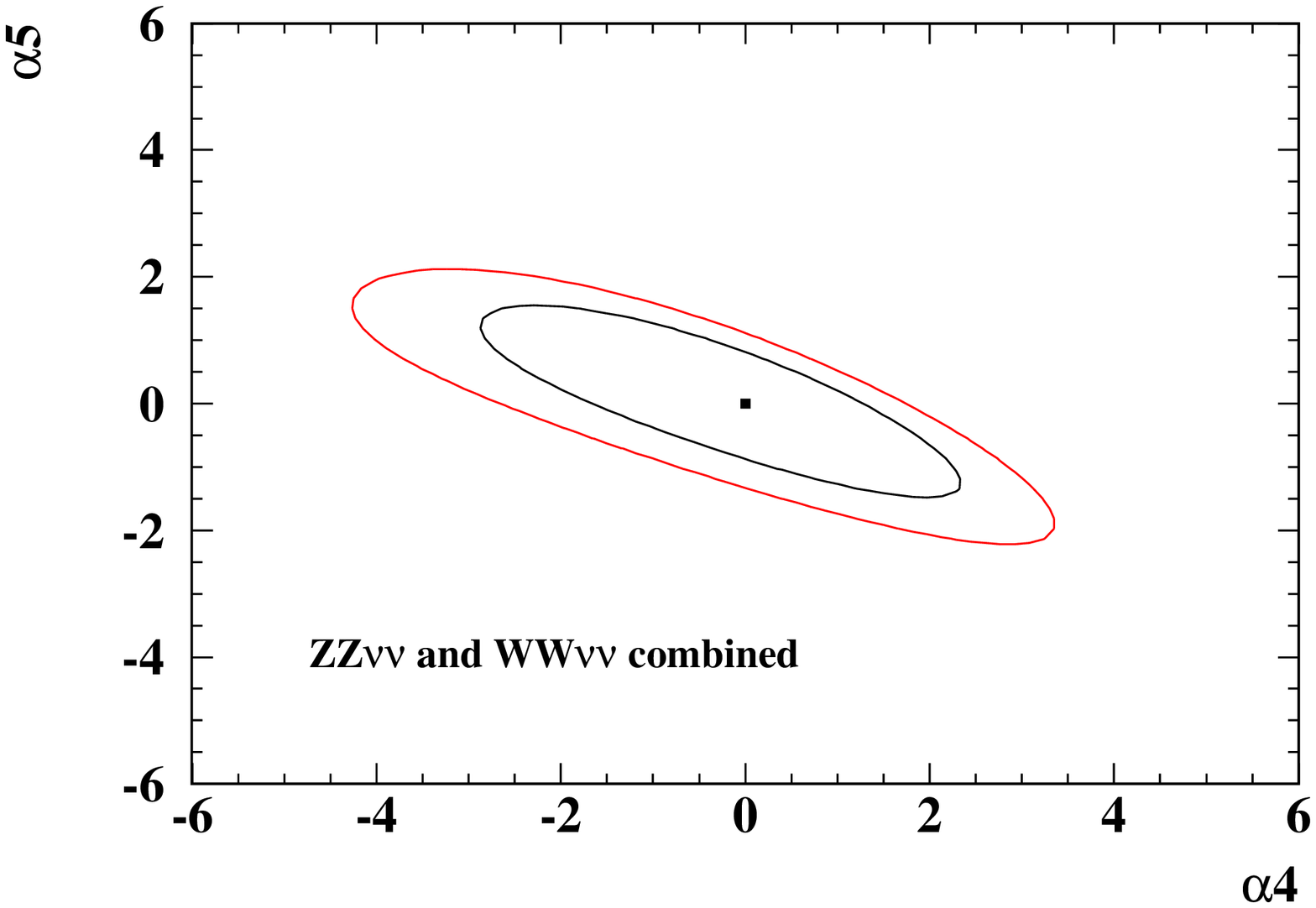}
\end{center}
\caption{
Sensitivity to $\alpha_4$ and $\alpha_5$ at $\sqrt{s}=800\GeV$ from
$\ee \rightarrow \nu \nu \WW$ and $\ee \rightarrow \nu \nu
\rm{ZZ}$. The inner and outer contours represent 68\% and 90\% c.l.
(${\cal L} = 1000 \fbi,\, \pmi=0.8,\,\ppl=0.4$).
}
\label{fig:alpha45}
\end{figure}

The LHC can see resonances coupling to W-pairs up to masses of 
about 2 TeV \cite{lhcprec}.
If these resonances are vector-like a linear collider can easily 
measure their couplings in W-pair production (see e.g. fig. \ref{fig:techni}).
If they are scalar or tensor particles some 
information on their couplings can be obtained from the measurement of
$\alpha_4$ and $\alpha_5$. However if the electroweak symmetry is really broken
in a strongly interacting scenario finally a linear collider with few TeV
centre of mass energy will be needed.

\section{Conclusions}
With a linear collider our knowledge of the gauge boson properties can
be increased largely in many respects. Running on the Z-pole and
the W-threshold allows significant progress in the precision of the
observables that are most sensitive to electroweak loop corrections:
the fermionic couplings of the Z and the mass of the W. These
measurements test the consistency of the then-Standard Model on the 
$10^{-4}$ level or allow to estimate still unknown parameters for
example in Supersymmetry.
Also some progress in the measurement of the CKM matrix in W-decays is
possible, helping in the understanding of CP-violation.

The most important progress, however, can be reached in the
measurement of interactions amongst gauge bosons. 
The gauge interactions are a necessary feature of a non Abelian gauge
group and the structure of triple gauge couplings can be checked with a
precision of few$\times 10^{-4}$. If a light Higgs exists 
a completely new class of precision observables sensitive to loop corrections
is accessible.
The gauge coupling measurements, however, are especially interesting
if no light Higgs exists.
With the triple boson couplings one
is sensitive to new physics that has to be present at the TeV scale to
avoid violations of unitarity at these energies.
In addition one can measure for the first time the scattering of
gauge bosons.
Since this
process is the one that violates unitarity the earliest it is the most
obvious place to look for strong electroweak symmetry breaking
effects. The disadvantage of this process is that the effective WW
centre of mass energy is significantly lower than the $\ee$ energy,
but at $\sqrt{s}=800 \GeV$ one should already be able to see effects
from new physics if no light Higgs exists, although with smaller
sensitivity than with the triple boson couplings.
At higher energies this process will become ideal to see new resonances.

In summary, gauge boson physics is an essential part of the linear collider
physics program.
In a scenario with a light Higgs it will be
indispensable to consolidate the model and to change its status from a
working hypotheses to a physics theory.
If no Higgs is seen the measurement of three and four boson interactions
will open up a window to the physics of electroweak symmetry breaking
which might not be directly visible at any next generation collider.

%
\begingroup\raggedright\endgroup



\end{document}